\documentclass[prb,amsmath,amssymb,twocolumn,lengthcheck,showpacs,floatfix,groupedaddress,superscriptaddress]{revtex4}
\usepackage{graphicx}
\usepackage[english]{babel}
\usepackage{times}
\usepackage{dcolumn}
\usepackage[usenames,dvipsnames]{color}
\usepackage{units}
\usepackage{bm}

\begin{document}

\title{Fine structure of spectra in the antiferromagnetic phase of
the Kondo lattice model}

\author{\v{Z}iga Osolin}

\affiliation{Jo\v{z}ef Stefan Institute, Jamova 39, SI-1000 Ljubljana,
Slovenia}

\author{Thomas Pruschke}

\affiliation{Institute for Theoretical Physics, University of G\"ottingen,
Friedrich-Hund-Platz 1, D-37077 G\"ottingen, Germany}

\author{Rok \v{Z}itko}

\affiliation{Jo\v{z}ef Stefan Institute, Jamova 39, SI-1000 Ljubljana, Slovenia}
\affiliation{Faculty  of Mathematics and Physics, University of Ljubljana, 
Jadranska 19, SI-1000 Ljubljana, Slovenia}

\date{\today}

\begin{abstract}
We study the antiferromagnetic phase of the Kondo lattice model on
bipartite lattices at half-filling using the dynamical mean-field
theory with numerical renormalization group as the impurity solver,
focusing on the detailed structure of the spectral function,
self-energy, and optical conductivity. We discuss the deviations from
the simple hybridization picture, which adequately describes the
overall band structure of the system (four quasiparticle branches in
the reduced Brillouin zone), but neglects all effects of the
inelastic-scattering processes. These lead to additional structure
inside the bands, in particular asymmetric resonances or dips that
become more pronounced in the strong-coupling regime close to the
antiferromagnet-paramagnetic Kondo insulator quantum phase transition.
These features, which we name ``spin resonances'', appear generically
in all models where the $f$-orbital electrons are itinerant (large
Fermi surface) and there is N\'eel antiferromagnetic order (staggered
magnetization), such as periodic Anderson model and Kondo lattice
model with antiferromagneitc Kondo coupling, but are absent in
antiferromagnetic phases with localized $f$-orbital electrons (small
Fermi surface), such as the Kondo lattice model with ferromagnetic
Kondo coupling. We show that with increasing temperature and external
magnetic-field the spin resonances become suppressed at the same time
as the staggered magnetization is reduced. Optical conductivity
$\sigma(\Omega)$ has a threshold associated with the indirect gap,
followed by a plateau of low conductivity and the main peak associated
with the direct gap, while the spin resonances are reflected as a
secondary peak or a hump close to the main optical peak. This work
demonstrates the utility of high-spectral-resolution impurity solvers
to study the dynamical properties of strongly correlated fermion
systems.
\end{abstract}

\pacs{71.27.+a, 72.15.Qm, 75.20.Hr, 75.30.Mb}

\maketitle

\newcommand{\vc}[1]{{\mathbf{#1}}}
\renewcommand{\Im}{\mathrm{Im}}
\renewcommand{\Re}{\mathrm{Re}}

\section{Introduction}

Heavy-fermion lanthanide and actinide materials have unusual
properties which still lack a complete microscopic understanding
despite many decades of continuous research \cite{hewson1997kondo,
RevModPhys.56.755}. Their name originates from high effective mass
enhancement of their Fermi-liquid quasiparticles and they are
noteworthy for phenomena such as unconventional (spin-mediated)
superconductivity, complex magnetism, huge thermopower, and, in
general, very rich phase diagrams. In some cases, these materials have
semiconducting or insulating properties at low temperatures (Kondo
insulators). Some well known heavy-fermion compounds are
Ce$_3$Bi$_4$Pt$_3$, YbB$_{12}$, CeNiSn, SmB$_6$, and CeRh$_2$Si$_2$
\cite{jaime2000closing,sugiyama1988field,mason1992,
batkova2006,PhysRevB.81.094403,degiorgi1996}.

The minimal model for this class of systems, the Kondo lattice model
(KLM), qualitatively describes many crucial features of real materials
in the low-temperature limit. It consists of a lattice of local
moments (representing the $4f$ or $5f$ orbitals) coupled to the
conductance-band electrons ($spd$ bands) through the on-site exchange
coupling $J$. At high temperatures, the $f$ electrons act as nearly
free spins, while the itinerant electrons are effectivelly decoupled,
thus the material behaves as a conventional metal. Upon cooling,
however, the itinerant electrons tend to screen the localized moments,
a process known in quantum impurity physics as the Kondo effect
\cite{RevModPhys.47.773}. The lattice version of the Kondo effect
leads to a coherent state which is a strongly renormalized Fermi
liquid with the $f$ states included in the Fermi volume (``large Fermi
surface'' ground state). 
This can also
be viewed as the hybridization between the conduction band and the
now itinerant $f$ levels. Exactly at half-filling, the chemical
potential lies inside the gap between the resulting effective bands
and the system is insulating, while at finite doping the chemical
potential lies in a part of the band with very flat dispersion, giving rise
to the heavy-fermion behavior. The hybridization manifests as a
resonance in the self-energy function,
$\Sigma(z)=\tilde{V}^2/(z-\tilde{\epsilon}_f)$, where $\tilde{V}$ is
the renormalized hybridization and $\tilde{\epsilon}_f$ the
renormalized $f$-orbital energy.

Despite its simplicity, the KLM has a complex phase diagram which is
not fully unraveled yet even within approximate approaches such as the
dynamical mean-field theory (DMFT) at the single-site level. On a
bipartite lattice at half-filling (that is, for exactly one
conductance-band electron per lattice site), the system is an
antiferromagnet for $J<J_c$ and a paramagnet for $J>J_c$; in both
cases it is an insulator. This quantum phase transition results from
the competition between the lattice RKKY interaction and the Kondo
effect \cite{doniach1977kondo}. In the antiferromagnetic DMFT solution
the local moments are itinerant for all values of $J$ and they never
decouple from the conduction band (i.e., there is no
itinerant-localized transition when the system turns
antiferromagnetic) \cite{hoshino2010}, thus Kondo physics actually
plays an important role throughout the antiferromagnetic (AFM) phase,
too. This is revealed by the fact that in the AFM state the
hybridization resonance in the self-energy function persists; it
simply becomes shifted in a spin- and sublattice-dependent manner as
the staggered magnetization is established.

In this work, we study the cross-over from the weak-coupling
(band/Slater antiferromagnet) to the strong-coupling (Kondo
antiferromagnet) regime, focusing on the detailed structure of the
self-energy function $\Sigma(z)$ and other dynamic quantities. We find
an interesting fine structure in the spectral functions revealed by
accurate numerical renormalization group (NRG) calculations. In the
momentum-resolved spectral functions we observe that the hybridized
bands are not truly degenerate at the band center and that the local
(momentum-integrated) spectral function exhibits narrow features
(``spin resonances'') inside the bands. They become more pronounced in
the strong-coupling Kondo antiferromagnet, where they can be easily
distinguished from the gap edges. The spin resonances are universal:
they appear for different lattice densities of states (Gaussian, Bethe
lattice, 2D and 3D cubic), in high-spin extensions of the KLM and in
the periodic Anderson model (PAM), which is another paradigmatic model
for heavy-fermion compounds. They decrease in amplitude as the
temperature is increased and disappear at the thermal AFM-PM phase
transition, thus they are a direct manifestation of the staggered
magnetization in the system. They are observable in optical
conductivity as a high-frequency hump or even as a distinct peak in
some parameter ranges. Their origin can be explained as follows.

The exchange coupling to the $f$ levels not only induces hybridization
(which is a coherent effect), but also leads to some incoherent
scattering at finite excitation energies, as described by the non-zero
imaginary part of the self-energy. This generically leads to
additional spectral structure within the DMFT self-consistency loop,
in particular to the ``spin resonances''. The self-energy function can
be modeled using an Ansatz with a single dominant hybridization pole
which is shifted somewhat away from the real axis; there is also some
additional fine structure the details of which are, however, not
crucial for the emergence of the spin resonances. 

This paper is organized as follows. In Sec.~\ref{mm} we describe the
Kondo lattice model and the DMFT(NRG) method. In Sec.~\ref{struc} we
study the analytical properties of the DMFT equations for the
long-range-ordered antiferromagnetic phase with A/B sublattice
structure (N\'eel state) and the typical spectral features resulting
from a phenomenological Ansatz for the self-energy functions featuring
a single hybridization resonance with spin-dependent parameters. This
is followed in Sec.~\ref{res} by the presentation of the results of
numerical DMFT calculations, including the dependence of the spin
resonance on the Kondo coupling, temperature and external magnetic field.
In Sec.~\ref{disc} we then discuss fits of $\Sigma(\omega)$ to the
Ansatz functions and provide the interpretation in terms of elastic
and inelastic scattering.

\section{Model and method}
\label{mm}

The simplest model that describes the basic physics of heavy fermions
is the Kondo lattice model\cite{doniach1977kondo,2006cond.mat.12006C}.
It consists of a non-interacting $spd$ band that is coupled
at each lattice site to an immobile quantum-mechanical spin
representing an $f$-orbital electron. The corresponding Hamiltonian is
\begin{equation}
\begin{split}
H &= \sum_{k\sigma} \epsilon_k c_{\sigma}^\dagger c_{\sigma} + J \sum_i
\bm{S}_i \cdot{} \bm{s}_i \\
& + \sum_i \left( g_c \mu_B B s_{i,x} + g_f \mu_B B S_{i,x} \right).
\end{split}
\end{equation}
Here $\epsilon_k$ is the dispersion relation for the non-interacting
conductance-band ($c$) states, 
\begin{equation}
\bm{s}_i = \sum_{\sigma \sigma'} c_{i\sigma}^\dagger 
\left( \frac{1}{2} \boldsymbol{\tau}_{\sigma \sigma'} \right) c_{i\sigma'}
\end{equation}
(with $\bm{\tau}$ as Pauli matrices) is the spin of the itinerant
electron at site $i$, $\vc{S}_i$ is the quantum-mechanical spin operator of
the localized $f$ moment ($S=1/2$, unless noted otherwise), and $J>0$
is the antiferromagnetic Kondo exchange coupling. We choose the
magnetic field to be oriented along the $x$-axis, since at weak fields
it generates a transverse staggered magnetization which we orient
along the $z$ axis; $g_c$ and $g_f$ are the $g$-factors, $\mu_B$ is
the Bohr magneton.

Most results shown are computed for the Bethe lattice which has
semicircular density of states
\begin{equation}
\rho_0(\epsilon) = \frac{2}{\pi D} \sqrt{1 - (\epsilon/D)^2},
\end{equation}
where $D$ is the half-bandwidth. Different lattice types are
considered in Sec.~\ref{univ}. We focus on the half-filling case,
$\langle n \rangle = 1$.

We study the lattice model with the dynamical mean-field theory
\cite{RevModPhys.68.13}, an approximation consisting in taking the
self-energy to be local, $\Sigma(\vc{k}, \omega) \rightarrow
\Sigma(\omega)$. The KLM is then mapped to an effective impurity
problem subject to self-consistency equations that relate the impurity
bath hybridization function to the self-energy. We iteratively solve
the impurity problem using the NRG method
\cite{RevModPhys.47.773,PhysRevB.21.1003,RevModPhys.80.395}. The
spectral functions are computed using the full-density-matrix NRG
approach \cite{PhysRevB.74.245114,PhysRevLett.99.076402} with a
discretization scheme that allows for improved spectral resolution at
high energy scales \cite{PhysRevB.79.085106}. Compared to prior
DMFT(NRG) works \cite{bodensiek2011low,PhysRevB.76.245101}, our
calculations are performed with significantly reduced spectral
broadening, thus sharp features away from Fermi level are much better
resolved. We use NRG discretization parameter $\Lambda=2$ with $N_z=8$
interpenetrating meshes for the $z$-averaging \cite{PhysRevB.41.9403}
and the method to directly calculate the self-energy introduced by 
R.~Bulla et al. \cite{bulla1998numerical} The broadening parameter was
$\alpha=0.25$ in most calculations \cite{weichselbaum2007}. To allow
for the antiferromagnetic phase, a bipartite lattice ($\alpha \in
\{A,B\}$) is used. In the presence of the magnetic field, at each DMFT
iteration we perform two NRG calculations, one for each sublattice:
the two subproblems are independent, because this is still a
single-site DMFT approximation with no inter-sublattice correlations.
Both sublattice self-energies then enter the DMFT self-consistency
equations. We assume no spin symmetry, as we have to allow for
components of the magnetization that are perpendicular to the magnetic
field. All quantities (spectral functions, self-energies) become full
$2 \times 2$ matrices and the NRG calculations become numerically
demanding and time consuming (see Appendix for the derivation and more
details). We stop the DMFT iteration once the absolute integrated
difference between local lattice Green's function become smaller than
$10^{-4}$.  In the absence of external magnetic field, the DMFT
iteration to the AFM solution is rapid and no Broyden mixing is
necessary to stabilize it (although it accelerates the convergence)
\cite{broyden}. When Broyden mixing is used, it is necessary to shift
the initial density of states in a spin-dependent way for the two
sublattices in order to induce the symmetry breaking to the AFM
phase. Some minimal shift is necessary, otherwise Broyden solver
converges to a metastable solution which is nearly paramagnetic with
some spin-dependent artifacts.

At finite fields, clipping of the self-energy matrix elements has been
employed to preserve the causality in the DMFT loop. We clip only the
imaginary part of the $\Sigma$. First, the diagonals are clipped to
\begin{equation}
\mathrm{Im} \Sigma_{\sigma \sigma}(\omega) < - \delta,
\end{equation}
where $\delta$ is a clipping parameter, typically chosen to be
$10^{-4}$ or less. This is a standard clipping procedure used also in
the case when $\Sigma$ is spin diagonal. In the next step, the
out-of-diagonal parts are clipped by the requirement
\begin{equation}
| \mathrm{Im} \Sigma_{\sigma \bar{\sigma} } | < \sqrt{ \mathrm{Im}
\Sigma_{\uparrow \uparrow} \mathrm{Im} \Sigma_{\downarrow \downarrow} },
\end{equation}
which ensures that the matrix $\mathrm{Im} \Sigma$ is negative
definite and therefore the self-energy causal.

There are many numerical methods which can reliably determine the
phase boundaries and various static quantities, but much less is known
about the dynamic quantities, such as the spectral functions. The most
widely used DMFT(QMC) technique (i.e., the DMFT using Quantum Monte
Carlo as the impurity solver) is formulated on the imaginary frequency
axis and requires resorting to an ill-posed analytical continuation to
obtain the final real-frequency results. This leads to uncertainties
and difficulties in resolving fine details in the spectral functions.
For example, to the best of our knowledge, the spin-polaron structure
of the Hubbard model has not yet been obtained using the DMFT(QMC),
but is resolved nicely with exact diagonalization or high-resolution
NRG as the impurity solver \cite{PhysRevB.85.085124}. Detailed
features in optical conductivity are also very difficult to obtain
using analytical continuation.

In the paramagnetic phase, we formulate the DMFT self-consistency loop
for the Kondo lattice by taking as the basic unit one conduction-band
$c$ site and one spin. The effective quantum impurity model then takes
the form of an Anderson impurity model (in the limit of no
interaction, $U \to 0$) with a side-coupled spin. In this case, the
self-energy function for the $c$ site, which we denote $\Sigma$, fully
describes the effect of the exchange coupling with the local moment.
This is not the only possible DMFT mapping: there is another
representation where the impurity model is the Kondo model. The
advantage of our approach is that the self-energy can be easily
computed in the NRG approach using the self-energy trick, and that the
DMFT self-consistency equation takes a simple form (the same as for
the Hubbard model):
\begin{equation}
\Delta(z) = z+\mu-\left[ G_{\mathrm{loc}}^{-1}(z)+\Sigma(z) \right],
\end{equation}
where $\Delta(z)$ is the hybridization function used as the input to the
impurity solver, $\mu$ is the chemical potential, and
$G_{\mathrm{loc}}^{-1}$ is the local ($\vc{k}$-averaged) lattice
Green's function, defined through
\begin{equation}
\begin{split}
G_\mathrm{loc}(z) &= \frac{1}{N} \sum_{\vc{k}}
\frac{1}{z+\mu-\epsilon_{\vc{k}}-\Sigma(z)} \\
&= \int
\frac{\rho_0(\epsilon)\mathrm{d}\epsilon}{[z+\mu-\Sigma(z)]-\epsilon} \\
&= G_0[z+\mu-\Sigma(z)].
\end{split}
\end{equation}
Here $\rho_0(\epsilon)$ is the dispersion relation for the
non-interacting conduction band, while $G_0(z)$ is the corresponding
non-interacting Green's function. Solving the impurity problem with
the NRG is equally costly for both possible DMFT mappings.

In the paramagnetic Kondo insulator at half-filling, the dominant
feature in the self-energy function is a pole \cite{hewson}:
\begin{equation}
\label{eq1}
\Sigma(z) = \frac{{\tilde{V}}^2}{z},
\end{equation}
where $\tilde{V}$ can be interpreted as the renormalized hybridization
in the hybridization picture. This generates an indirect spectral gap
approximately given by
\begin{equation}
\Delta \approx \frac{{\tilde V}^2}{D}.
\end{equation}
The optical conductivity $\sigma(\Omega)$ remains low at frequencies
above $2\Delta$ until the main peak occurs for $\Omega \approx
2\omega^*$, where $\omega^*$ is the direct gap that corresponds to the
frequency where the quasiparticle band in the momentum-resolved
spectral function crosses the $\epsilon_\vc{k}=0$ line
\cite{rozenberg1996}. It is thus defined through
\begin{equation}
\omega^* = \Re\Sigma(\omega^*),
\end{equation}
which then leads to 
\begin{equation}
\omega^* \approx \tilde{V}.
\end{equation}
A quantity defined in a similar way as $\omega^*$ will play a
prominent role in the antiferromagnetic case, too.

While Eq.~\eqref{eq1} is an excellent approximation, in reality
$\Sigma(z)$ has some additional features in the energy range outside
the gap  and $\mathrm{Im}\Sigma$ is non-zero except inside the gap.

\section{Analytical properties of DMFT equations for bipartite lattices}
\label{struc}

In its simplest form, the DMFT approach is applied to homogeneous
phases where all lattice sites are equivalent, $\Sigma_i=\Sigma$, but
it can also be used to study phases with commensurate
antiferromagnetic long-range order \cite{georges1996}. In a bipartite
lattice, for example, N\'eel order can be described by two different
self-energy functions, $\Sigma_A$ and $\Sigma_B$, for lattice sites
belonging to either sublattice.

Let us first consider the case without external magnetic field.
Working in the reduced Brillouin zone for an enlarged unit cell
consisting of one A site and one B site, the band Hamiltonian is
\begin{equation}
H_0 = \sum_{\sigma,\vc{k} \in \mathrm{RBZ} } \epsilon_\vc{k} \left(
c^\dag_{A\vc{k}\sigma} c_{B\vc{k}\sigma} + \text{H.c.} \right),
\end{equation}
and the inverse lattice Green's function matrix is
\begin{equation}
G_{\vc{k}\sigma}^{-1}(z) = \begin{pmatrix}
z +\mu-\Sigma_{A\sigma}(z) & -\epsilon_{\vc{k}} \\
-\epsilon_{\vc{k}} & z+\mu-\Sigma_{B\sigma}(z)
\end{pmatrix},
\end{equation}
thus
\begin{equation}
G_{\vc{k}\sigma}(z) =
\frac{1}{\zeta_{A\sigma}\zeta_{B\sigma}-\epsilon_\vc{k}^2}
\begin{pmatrix} 
\zeta_{B\sigma} & -\epsilon_{\vc{k}} \\
-\epsilon_{\vc{k}} & \zeta_{A\sigma} 
\end{pmatrix},
\end{equation}
where $\zeta_{\alpha\sigma}(z)=z+\mu-\Sigma_{\alpha\sigma}(z)$, $\alpha =
A,B$. The local Green's functions are then obtained through
integration. The out-of-diagonal elements are zero due to the
particle-hole symmetry of the band (this is the case also at finite
doping). The diagonal Green's functions are given by
\begin{equation}
\label{eqloc}
G_{\mathrm{loc},\alpha\sigma}(z) = \zeta_{\bar{\alpha}\sigma}(z)
\int
\frac{\rho_0(\epsilon)\mathrm{d}\epsilon}{\zeta_{A\sigma}(z)
\zeta_{B\sigma}(z)-\epsilon^2},
\end{equation}
and the spectral function is given through
\begin{equation}
\label{eqA}
A_{\alpha\sigma}(\omega) = -\frac{1}{\pi} \mathrm{Im}
G_{\mathrm{loc},\alpha\sigma}(\omega+i0^+).
\end{equation}
Using fraction expansion, the integrals can be expressed in closed
form. This gives
\begin{equation}
\label{eq22}
G_{\mathrm{loc},\alpha\sigma}(z) = \zeta_{\bar{\alpha}\sigma}
\frac{
G_0(\sqrt{\zeta_{A\sigma}\zeta_{B\sigma}})
-G_0(-\sqrt{\zeta_{A\sigma}\zeta_{B\sigma}})
}{2 \sqrt{\zeta_{A\sigma} \zeta_{B\sigma}}}.
\end{equation}
The DMFT loop is then closed via a site ($A/B$) and spin-dependent
self-consistency equation
\begin{equation}
\Delta_{\alpha\sigma}(z) = z+\mu-\left[
G_{\mathrm{loc},\alpha\sigma}^{-1}(z)+\Sigma_{\alpha\sigma}(z) 
\right].
\end{equation}
The hybridization matrix is diagonal. In the absence of external
magnetic field, one can thus use simple U(1)$_\mathrm{spin}$ NRG code
with spin-dependent Wilson chains. A further simplification in this
case is provided by the symmetry relations
$G_{A\sigma}=G_{B\bar{\sigma}}$ and
$\Sigma_{A\sigma}=\Sigma_{B\bar{\sigma}}$. In general, however, one
must use the full $2 \times 2$ matrix structure in the spin space and
properly handle the discretization of the hybridization matrix with
out-of-diagonal elements. The derivations and implementation details
are discussed in Appendix~\ref{appA}.

If the $f$ electrons remain itinerant in the AFM phase, as indicated
by the DMFT calculations, one expects that the hybridization picture
remains approximately correct in the ordered phase, but one needs to
incorporate the effects of the exchange fields in the lattice. Writing
these fields as $h_\alpha=\pm h$ for the $c$-band and $H_\alpha=\pm H$
for the $f$-band, respectively (plus sign for sublattice $A$, minus
sign for sublattice $B$), one then expects the following form of the
self-energy
\begin{equation}
\label{pole}
\Sigma_{\alpha\sigma}(z) = \alpha\sigma h + \frac{{\tilde
V}^2}{z-\alpha\sigma H},
\end{equation}
where on the right-hand side $\alpha$ and $\sigma$ are to be understood as
$\pm 1$ factors. This leads to excitation branches given by
\begin{equation}
E(\vc{k}) = \pm \frac{1}{2} \left[
(\epsilon_\vc{k}+h+H \pm \sqrt{(\epsilon_\vc{k}+h-H)^2+4{\tilde V}^2} 
\right].
\end{equation}
Previous work based on the continious-time QMC solver suggested that the
relation
\begin{equation}
\label{hh}
h=-H
\end{equation}
(named ``quasilocal compensation'') is satisfied \cite{hoshino2010}.
This constraint was said to results from Kondo physics and lattice
coherence, since the efective energy levels in the hybridization
picture for itinerant antiferromagnetism in the KLM are determined not
by local exchange fields, but by long-ranged molecular fields
involving distant conduction-band electrons \cite{hoshino2010}. In
case of perfect quasilocal compensation, the quasiparticle branches
intersect at $\epsilon_\vc{k}=0$ and the local spectral functions are
quite similar to those for the Kondo insulator, only with staggered
spin polarization.

\begin{figure*}[ht!] \centering
\includegraphics[clip,width=0.9\textwidth]{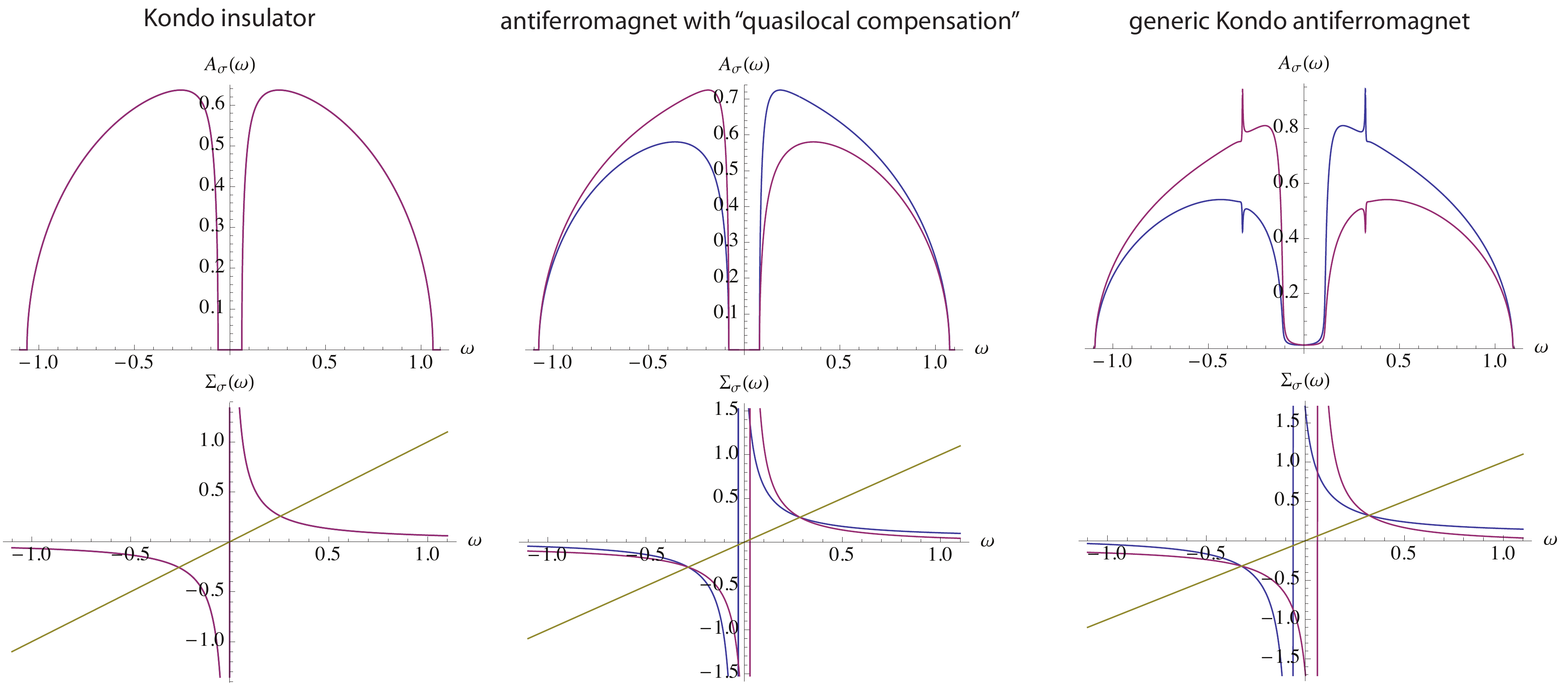}
\caption{(Color online) Local spectral functions $A_\sigma(\omega)$
(top row) computed for self-energies $\Sigma_\sigma(\omega)$ (bottom
row) approximated by a single pole on or near the real axis for
bipartite lattice that allows AFM order. In the Kondo insulator, there
is no spin splitting and the pole is located in the center of the gap
at $\omega=0$ on the real axis. In the antiferromagnetic state with
``quasilocal compensation'', the poles are located symmetrically on the
real axis at $\omega=\pm H$ and the real parts of $\Sigma$ are
additionally shifted by $\pm h=\mp H$. In generic Kondo
antiferromagnet, the values of $h$ and $-H$ are not exactly the same
and the poles are shifted away from the real axis (i.e., $\Im\Sigma$
is non-zero).}
\label{mma}
\end{figure*}

If the quasilocal compensation, Eq.~\eqref{hh}, is violated, there is
an avoided crossing between the quasiparticle branches. This should in
principle lead to an opening of additional gaps, however, since this
is a strongly interacting system, the self-energy has non-zero
imaginary part and the pole in Eq.~\eqref{pole} can lie away from the
real axis. This immediately implies that there will be some additional
structure inside the bands at energies $E=\pm\omega^*$ where
$\omega^*$ is now approximately (assuming $h\approx -H$)
\begin{equation}
\omega^* \approx \sqrt{h^2+{\tilde V}^2}.
\end{equation}
We show in the following that the combination of inelastic scattering
(broadening) and the avoided crossings of quasiparticle bands result
in asymmetric resonance curves in the local spectral functions (``spin
resonances''), as shown in the schematic plots in Fig.~\ref{mma}.

These analytical considerations thus suggest that fine structures are
expected quite generically in the DMFT solution. In previous
DMFT(QMC), they were not visible, presumably due to difficulties in
performing analytic continuations. As we show in the following
section, they can be resolved using the DMFT(NRG) approach.

\section{DMFT results}
\label{res}

\subsection{Spin resonance structures}

\begin{figure*}[ht!] \centering
\includegraphics[clip,width=0.9\textwidth]{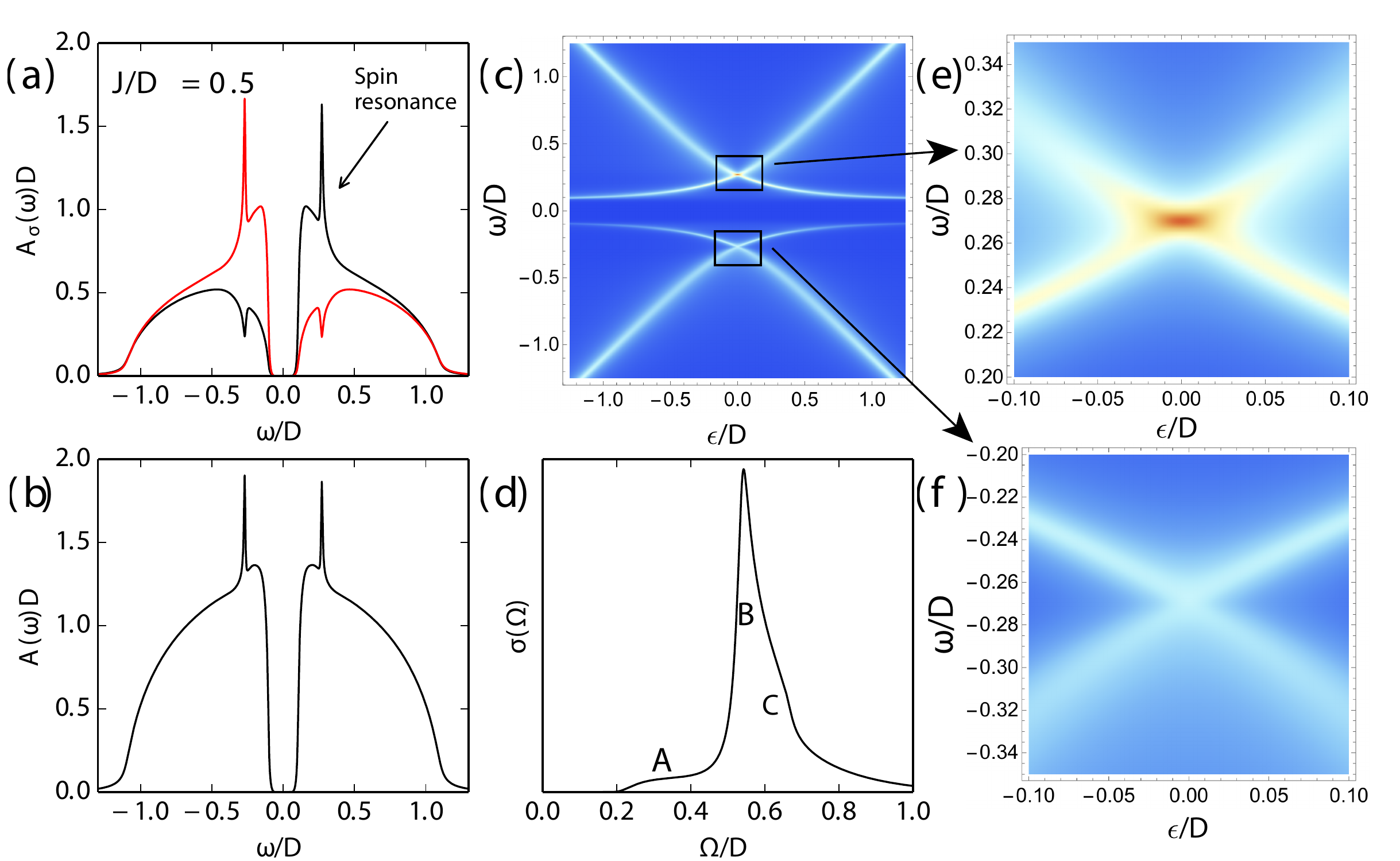}
\caption{(Color online) Multiple manifestations of the ``spin
  resonance'' (fine structure inside the bands) in the dynamical
  properties of the Kondo lattice model in the antiferromagnetic
  phase: (a) Peaks and dips in the spin-resolved local spectral
  function $A_\sigma(\omega)$ of the conduction band. (b) Peaks in the
  spin-averaged local spectral function
  $A(\omega)=A_\uparrow(\omega)+A_\downarrow(\omega)$. (c)
  Momentum-resolved spectral function $A(\epsilon,\omega)$ and
  close-ups on the regions at $\epsilon=0$ showing (e) an enhanced
  density of states in the empty band (associated with the resonance)
  and (f) a reduced density of states, i.e., avoided crossing, in the
  occupied band (associated with the dip). (d) Optical conductivity
  exhibiting a hump for energies slightly above the main maximum.
  Labels (A), (B), and (C) are discussed in the text.}
\label{fig1}
\end{figure*}

In Fig.~\ref{fig1} we summarize the main results of this work for a
value of $J$ in the parameter range where the effects are the most
pronounced, i.e., in the strong-coupling case near the AFM-KI
transition. In the spectral function of the $c$-band electrons, we
observe an additional structure inside the band,
Fig.~\ref{fig1}(a). In the occupied band, there is a dip for minority
spin and a sharp peak for majority spin; the resonance is also visible
in the spin-averaged spectral function, $A = A_{\downarrow} +
A_{\uparrow}$ shown in Fig.~\ref{fig1}(b).
The origin of these features can be traced back to the
behavior of the momentum-resolved spectral function
$A(\vc{k},\omega)$, plotted as a function of $\epsilon_\vc{k}$ in
Fig.~\ref{fig1}(c). The close-ups on the regions where the
quasiparticle branches should intersect reveal that the spectral dip is
associated with a reduced spectral weight between the branches, i.e.,
an avoided crossing, Fig.~\ref{fig1}(f), while the peak corresponds to
an enhancement between two branches, Fig.~\ref{fig1}(e). 

The optical conductance, $\sigma(\Omega)$ shown in Fig.~\ref{fig1}(d),
shows a threshold at twice the indirect gap $\Delta$, then remains
roughly constant up to a sizeable peak at $\Omega \approx 2\omega^*$,
where transitions between two pairs of bands are strongly enhanced due
to the cross-shaped momentum-resolved spectral functions for both
occupied and empty bands. The presence of the spin-resonance is
reflected in the shape of this peak which has a notable hump in its
high-frequency flank.

\begin{figure} \centering
\includegraphics[clip,width=0.48\textwidth]{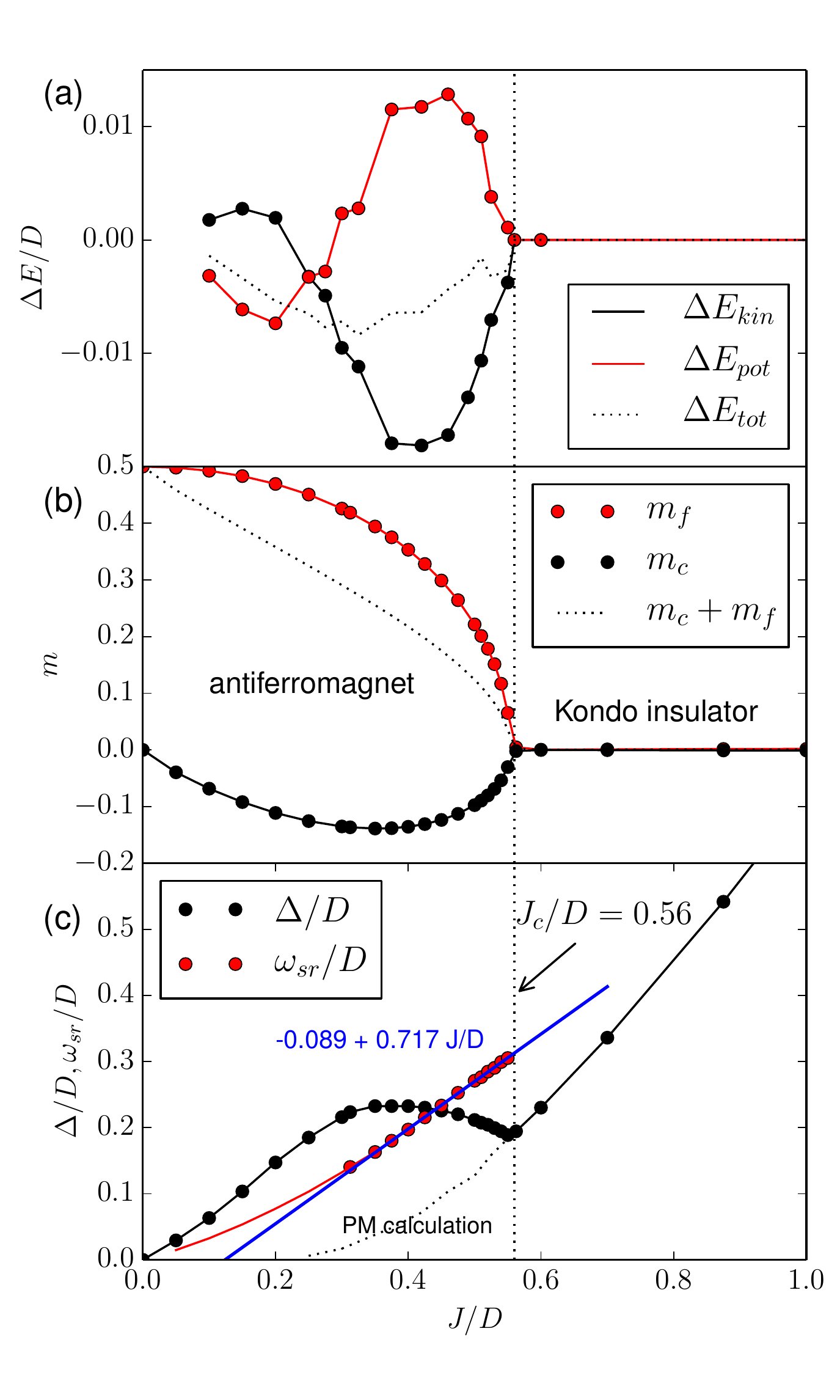}
\caption{(Color online) From weak-coupling to strong-coupling
antiferromagnetism in the Kondo lattice model as a function of the
Kondo exchange coupling $J$. a) Reduction of kinetic viz. potential
energy with respect to the reference paramagnetic state, indicating
different mechanisms of magnetic ordering for weak and strong coupling
regimes. b) Staggered magnetization of the conduction band electrons
($m_c$) and local moments ($m_f$). c) Spectral gap $\Delta$ and
spin-resonance position $\omega_{sr}$.
$\omega_{sr}$ is not well defined for $J/D \lesssim 0.3$.}
\label{fig2}
\end{figure}

\begin{figure} \centering
\includegraphics[clip,width=0.48\textwidth]{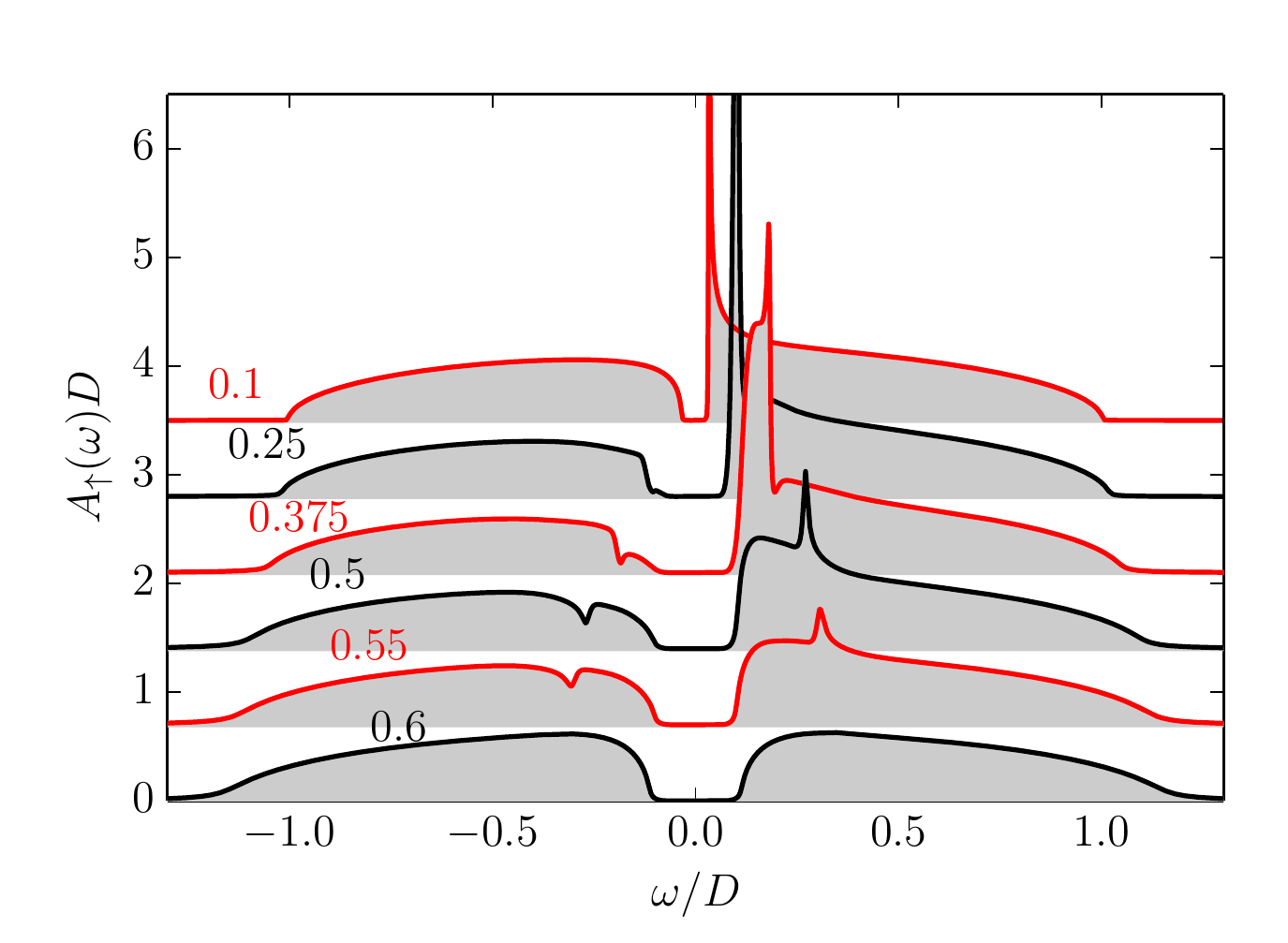}
\caption{(Color online) Spin-resolved spectral functions of the
conduction band for a range of $J$: (a) weak-coupling Slater regime
with inverse square root divergence at gap edges ($J/D=0.1$,
$0.25$), (b) cross-over regime with a complex form of the spectra near
the gap edges ($0.375$), (c) strong-coupling regime with smooth gap
edges and well-developped spin resonances ($0.5$, $0.55$), (d) Kondo
insulator with no spin resonance ($0.6$).}
\label{fig3}
\end{figure}

\subsection{Single-particle properties}

The origin of antiferromagnetism in the KLM depends on the value of
the exchange coupling $J$, see Fig.~\ref{fig2}(a). For small $J$, the
AFM order develops by a mechanism similar to the Slater
antiferromagnetism in the Hubbard model (although in the KLM, the
system would be insulating even in the absence of the unit cell
doubling and consequent gap opening due to magnetic order). This
regime can be fully explained within a simple Hartree-Fock (HF) theory
\cite{rozenberg1995}: the $f$ states are fully polarized, while the
$c$ states weakly anti-align with the $f$ spins at each site,
Fig.~\ref{fig2}(b). There are inverse square root Slater singularities
at the gap edges, Fig.~\ref{fig3} for $J/D=0.1$. The quasiparticle gap
is linear in $J$ due to the nesting instability
($\epsilon_{k_F+q}=-\epsilon_{-k_F+q}$) to AFM order in the p-h
symmetric case \cite{capponi2001}, leading to $\Delta =Jm_f$, where
$m_f$ is the staggered magnetization of the $f$ orbitals. The $f$
spins are nearly fully polarized ($m_f \to 1/2$ as $J \to 0$), while
$c$ electrons start out unpolarized ($m_c \to 0$ as $J \to 0$). 

In the intermediate regime, $J \sim J' \approx 0.4D$, the spin
resonance gradually develops from the band edge, Fig.~\ref{fig3}. In
this regime, the gap vs. $J$ curve flattens out to form a broad
plateau that peaks around $J/D=0.35$. The staggered magnetization of
the conduction-band electrons is also maximal in this parameter range.

For $J>J'$, in the strong-coupling regime, the AFM is driven by the
reduction in the kinetic energy and is characterized by a
well-resolved spin resonance, Fig.~\ref{fig3}(c). The staggered
magnetizations $m_c$ and $m_f$, as well as the gap $\Delta$ are all
decreasing in this regime. Finally, as $J$ is increased further, there
is a second-order quantum phase transition to a paramagnetic Kondo
insulator state, Fig.~\ref{fig3}(d). The charge gap is continuous
across the transition.

In the intermediate to strong-coupling regime, the band gap $\Delta$
is non-linear, even non-monotonous, function of $J$, while the spin
resonance position $\omega_{sr}$ behaves linearly, see
Fig.~\ref{fig2}(c). A good fit is given by
\begin{equation}
\omega_{sr} = 0.717 J - 0.089.
\end{equation}
This linearity is ``inherited'' from the Kondo insulator phase, where
it holds for the quantity $\omega^*$.
This further emphasizes the continuous nature of the AFM-KI
phase transition and the persistence of itinerancy.

\subsection{Optical conductivity}

\begin{figure} \centering
\includegraphics[clip,width=0.48\textwidth]{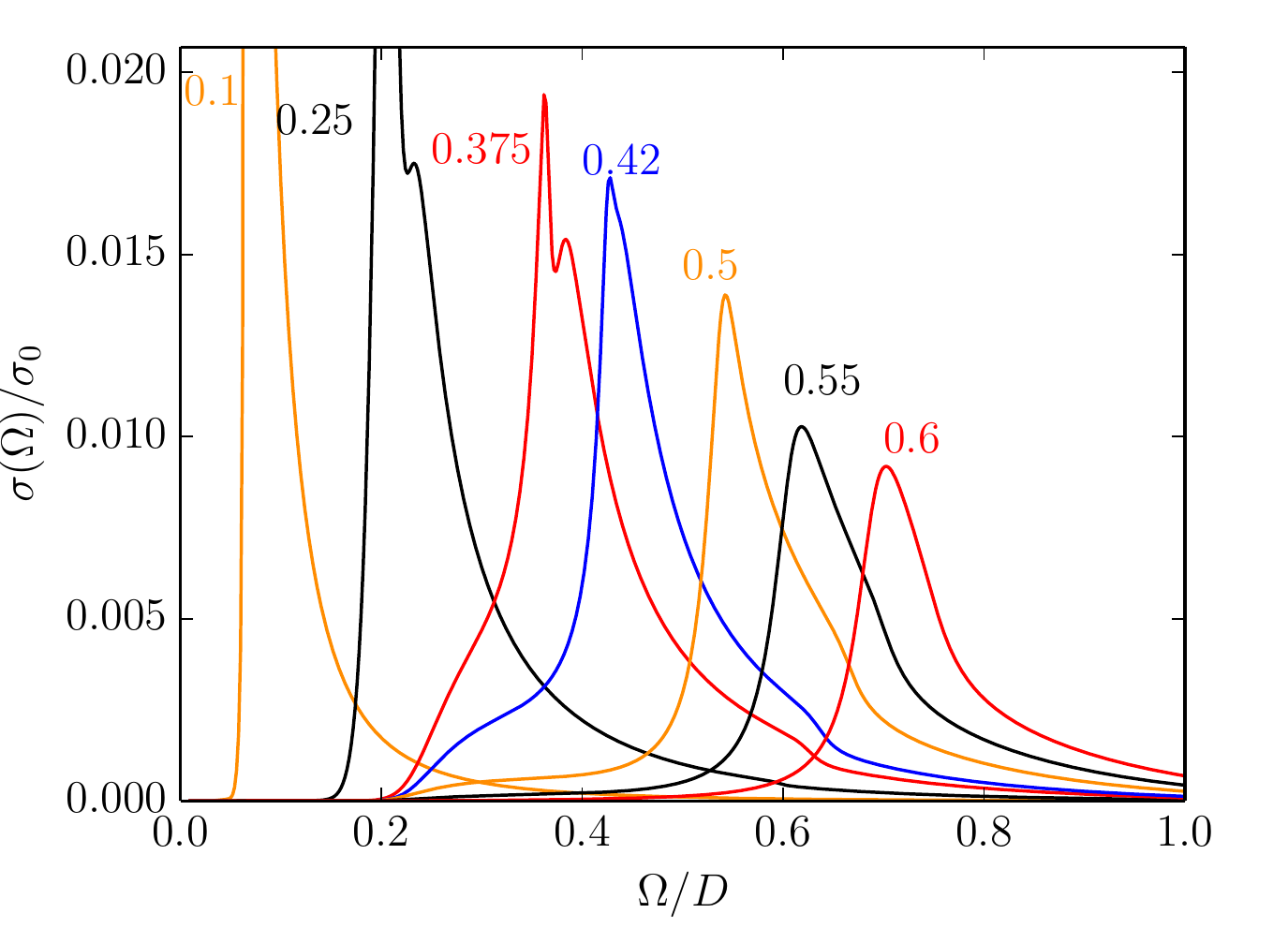}
\caption{(Color online) Optical conductivity for a range of $J/D$ (as
indicated in the plot next to the corresponding curves). Weak-coupling
antiferromagnetism is characterized by threshold behavior with a peak
at $2\Delta$, while strong-coupling antiferromagnetism exhibits more
complex behavior with threshold at $2\Delta$ and the main peak at
$2\omega^* \approx 2 \omega_\mathrm{sr}$, the spin resonance
appearing as a secondary maximum or as a hump on the flank of the main
peak. $\sigma_0=e^2/h$.}
\label{fig4}
\end{figure}

In the weak-coupling AFM regime, the optical conductivity shows a
threshold behavior with a pronounced resonance corresponding to twice
the quasiparticle gap, $\Omega=2\Delta$, see Fig.~\ref{fig4} for
$J/D=0.1$. Similar behavior is also observed in the Slater AFM regime
of the Hubbard model \cite{zitzler2002}. In the strong-coupling
regime, the curves are more complex, Fig.~\ref{fig1}(d). After the
threshold at $\Omega = 2\Delta$, there is (A) a region of moderate
conductivity, followed by (B) a pronounced resonance at $\Omega =
2\omega^*$, and (C) an additional more-or-less pronounced structure
associated with the spin-resonance. As $J$ is increased toward $J_c$,
region A progressively flattens out and evolves into a plateau of
nearly constant very low optical conductivity (see $J/D=0.375$,
$0.42$, and $0.50$ in Fig.~\ref{fig4}). This region is associated with
the transitions between the quasiparticles at band edges
($\epsilon_\vc{k} \approx \pm D$) which have low spectral weight.
Region B is associated with the cross-shaped form of the
$\epsilon_\vc{k}$-resolved spectral function in the band-center
($\epsilon_\vc{k}\approx0$).

\begin{figure} \centering
\includegraphics[clip,width=0.48\textwidth]{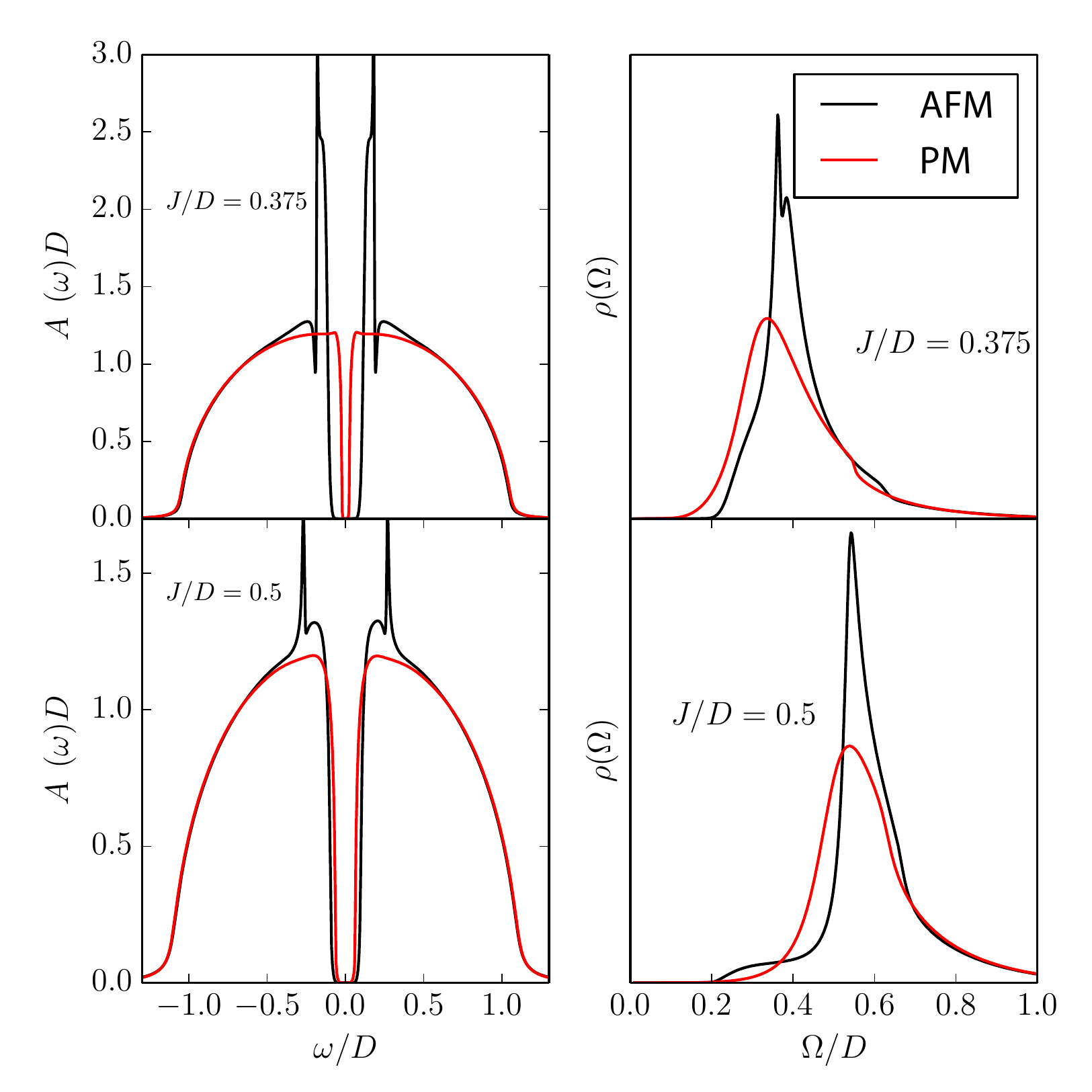}
\caption{(Color online) Comparison of antiferromagnetic and
paramagnetic DMFT solutions for equal values of $J$ reveals the
fine details in the ordered phase.}
\label{afmpm}
\end{figure}

It is worth to emphasize that the spin resonances are not observed for
negative (ferromagnetic) Kondo exchange coupling $J$, although the
system is also antiferromagnetic. This is due to the very different
topology of the quasiparticle bands (``small Fermi surface'') in this
case \cite{hoshino2010,oliver}, which is, in turn, associated with
a different form of the self-energy function with no pronounced poles.
Furthermore, there is no spin resonance if we enforce paramagnetic
solution in the region where the AFM is the true ground state (such a
comparison of AFM and PM solutions in shown in Fig.~\ref{afmpm}); in the
paramagnetic case the topology is that of ``large Fermi surface'', but
there is no staggered magnetization. We thus conclude that the {\sl spin
resonance is a characteristic property of itinerant
antiferromagnetism, requiring both itinerancy of $f$ electrons and
staggered magnetization}.

In DMFT calculations using solvers requiring an analytical
continuation, such in-band spectral features have not been observed.
This is the case also for high-quality continuous-time QMC
calculations \cite{hoshino2010}. Some hints of the spin resonances
have been observed in prior DMFT(NRG) works
\cite{peters2007magnetic,Peters:2011iq,oliver}, but have not been
discussed. The spin resonances appear for any value of the NRG
broadening parameter: even in calculation with no $z$-averaging and
with large broadening parameter their presence is suggested by a broad
spectral hump in one band and as a faint depression in the other. As
the broadening is decreased, these features become sharper and more
asymmetric. Because of their persistent nature and very generic
conditions on the functional form of the self-energy for their
emergence, it is unlikely that they were a numerical artifact of NRG
calculations.

\subsection{Temperature dependence}

\begin{figure} \centering
\includegraphics[clip,width=0.48\textwidth]{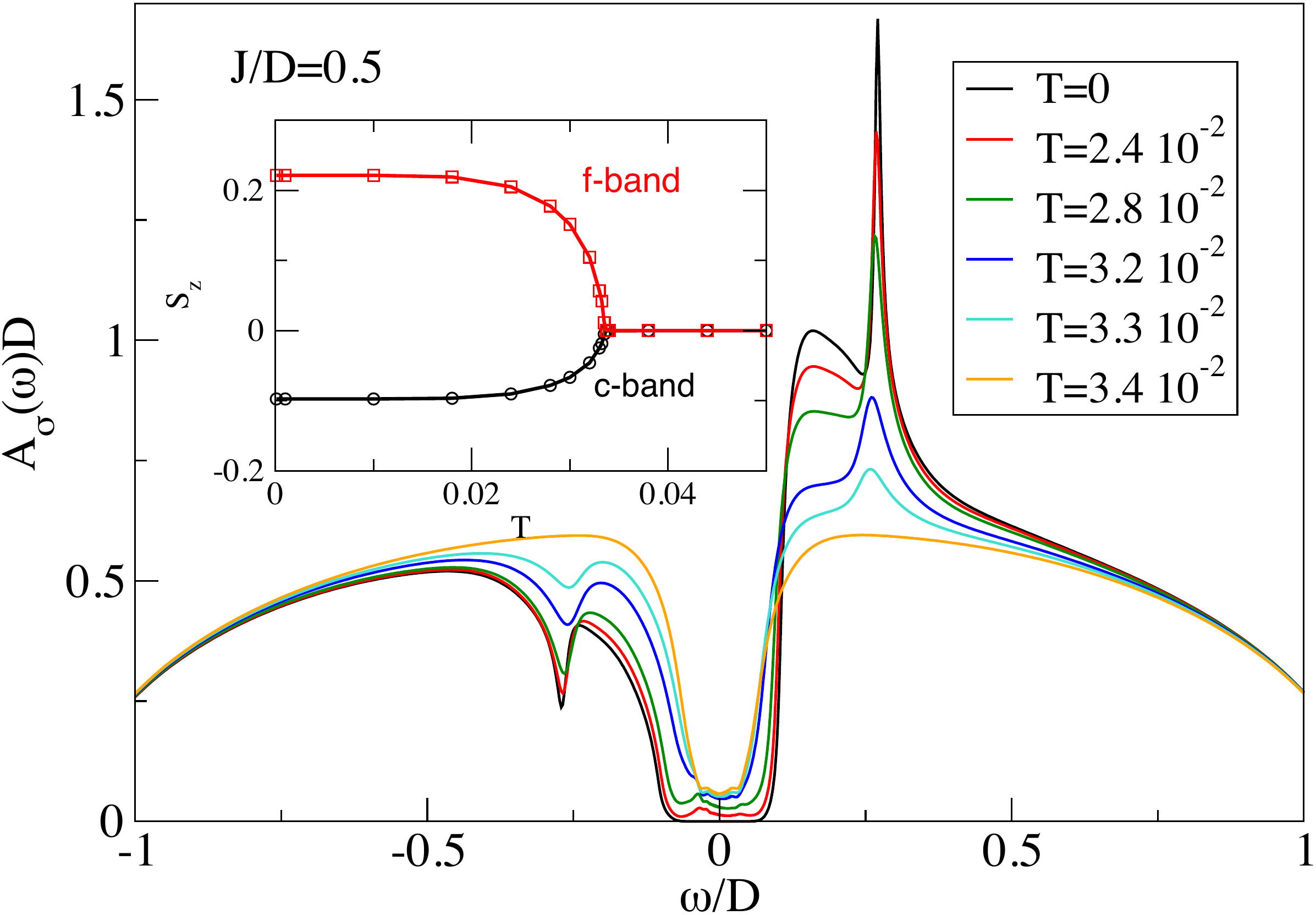}
\caption{(Color online) Reduction of the spin resonance structure
(peak weight) with increasing temperature. Inset: temperature
variation of the staggered magnetization across the thermal AFM-PM
phase transition.}
\label{fig5}
\end{figure}

The staggered magnetization decreases with increasing temperature
until at the N\'eel temperature $T_N$ the system undergoes a
transition to the paramagnetic phase, Fig.~\ref{fig5}. The evolution
of the spectra confirms the relation of the spin-resonance peaks with
the magnetic order, since the peak intensity follows the staggered
magnetization. Interestingly, the peak position itself does not depend
much on the order parameter.  It is also noteworthy that the overall
structure of the effective bands does not change accross the
transition \cite{hoshino2010}. A sign of this is the persistence of a
reduced density of states (a hybridization-induced ``pseudo-gap'')
around $\omega=0$ to temperatures well above $T_N$, where the order
parameter is already zero and the spin resonance structure eliminated.

\subsection{Magnetic field}

We now consider the effect of an external magnetic field on the
antiferromagnetic state. We assume $g \equiv g_c = g_f$ and express
the field in units of the Zeeman energy, $g\mu_B B$. There are no
magnetic anisotropy terms in our Hamiltonian, thus in the ground state
the staggered magnetization always reorients itself perpendicular to
the applied field to preserve the exchange energy generated by the
antialignment of spins in $c$ and $f$ bands. \footnote{If the external
magnetic field is applied in the direction of stagerred magnetization,
a metastable solution with no transverse magnetization can be obtained
in DMFT calculations. We do not consider this case here.} Likewise,
when magnetic field is applied on the Kondo insulator, it induces an
antiferromagnetic phase transverse to the external field
\cite{PhysRevLett.92.026401,PhysRevB.70.245104}. In this work we will
follow the convention that the direction of the field is taken to be
along the $x$ axis (we denote this as the ``longitudinal'' direction)
and the staggered magnetization along the $z$ axis (this is the
``transverse'' direction).

\begin{figure} \centering
\includegraphics[clip,width=0.48\textwidth]{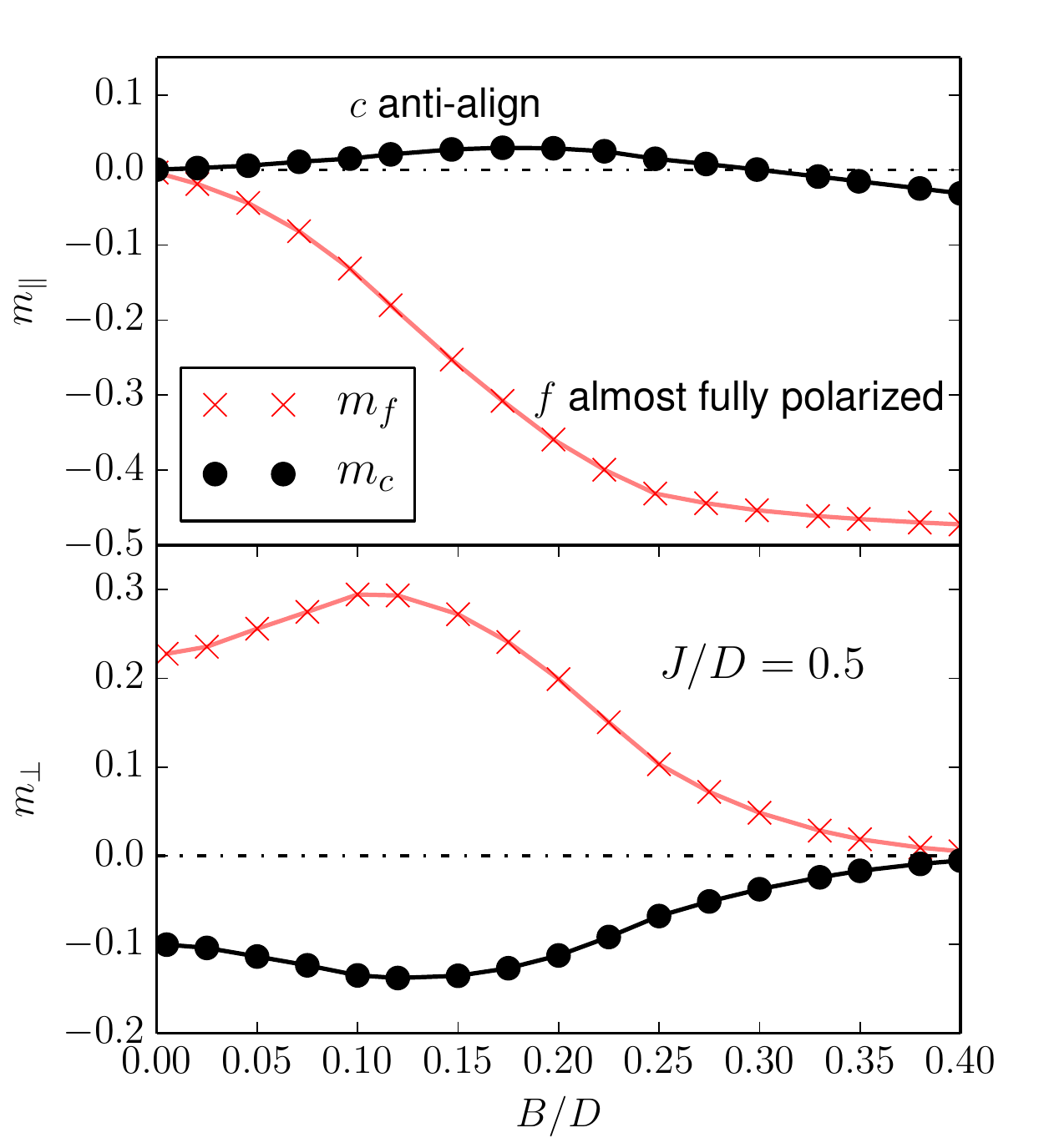}
\caption{(Color online) Effect of an applied magnetic field on the
magnetic order in the antiferromagnetic phase of the Kondo lattice
model. Upper panel: uniform magnetization components $m_\parallel$ of
$c$ and $f$ orbitals in the direction of the field. Bottom panel:
staggered magnetization components $m_\perp$ of $c$ and $f$ orbitals
perpendicular to the field. The following defining relations hold:
$m_{x,A}=m_{x,B}=m_{\parallel}$ and $m_{z,A} = -m_{z,B} = m_{\perp}$.
}
\label{fig7}
\end{figure}

$f$ electrons have higher magnetic susceptibility than $c$ electrons,
thus their uniform magnetization rapidly increases with the applied
field, while the $c$ electrons at first anti-align due to the strong
local Kondo coupling $J \gg B$ and only for very strong fields (of
order $J$) reorder in the same direction as the $f$ orbitals, see
upper panel in Fig.~\ref{fig7}. For weak fields, the stagerred
magnetization first increases, see lower panel in Fig.~\ref{fig7}.
This can be explained as the suppression of the Kondo effect by
breaking the local singlets through magnetic field, which leads to
stronger spin polarization of the orbitals. The staggered
magnetization is maximal for $B=B_m \approx 0.1D$
and then slowly decreases towards $0$ as the gap is closing. The
charge gap becomes exponentially small in the large-$B$ limit
\cite{PhysRevLett.92.026401}, thus at non-zero temperature the system is effectively a
strongly spin-polarized paramagnetic metal. The results in
Fig.~\ref{fig7} can be qualitatively reproduced using a simple exact
calculation on a two-site cluster with suitable molecular fields for
AFM order put in by hand. The only small discrepancies are due to the
stronger itinerancy of $c$ electrons in the full latice.

\begin{figure} \centering
\includegraphics[clip,width=0.48\textwidth]{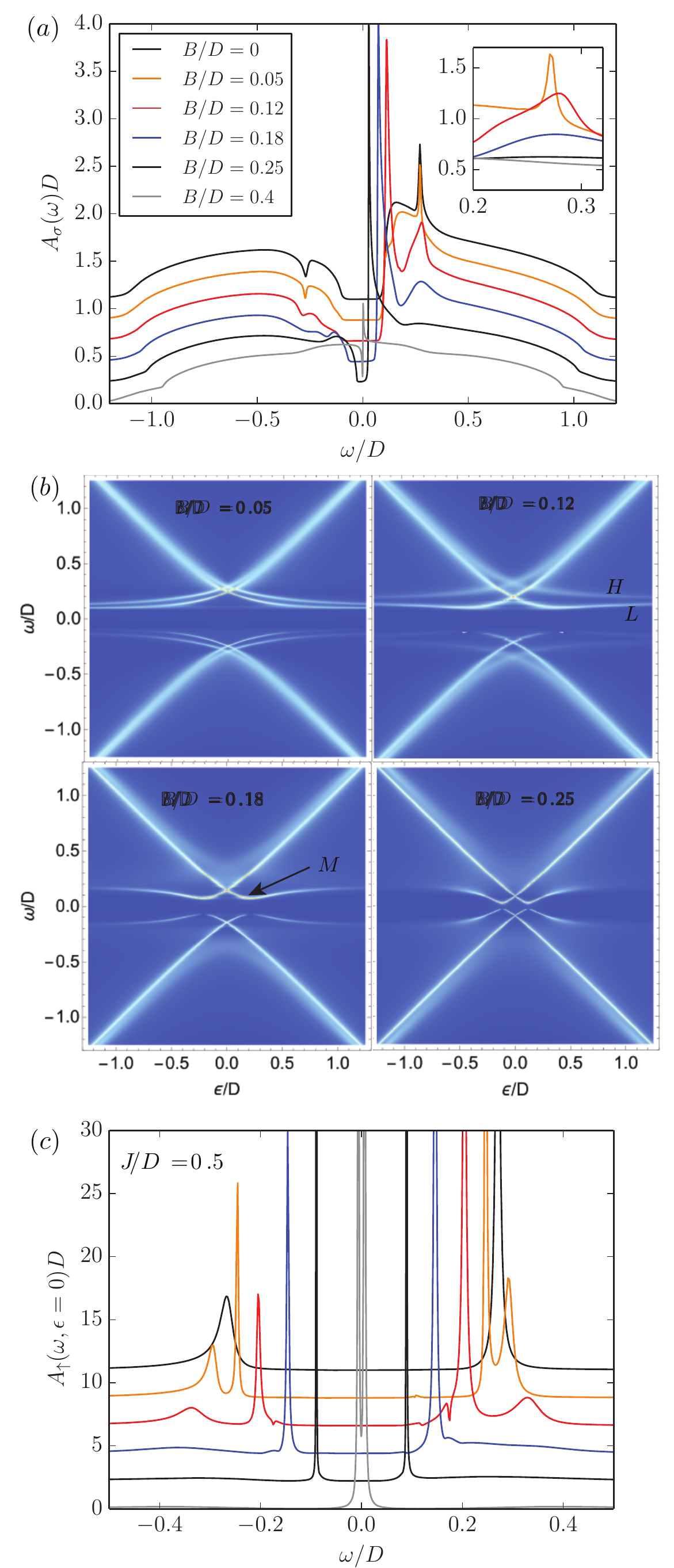}
\caption{(Color online) Evolution of (a) local spectral function
$A_\sigma(\omega)$ and (b) momentum-resolved spectra
$A_\sigma(\epsilon_\vc{k},\omega)$ for increasing external magnetic
field $B$ in the strong-coupling AFM phase ($J/D=0.5$).  Labels L, H,
and M are explained in the text. The cross-cut of the
momentum-resolved spectrum at $\epsilon=0$ is shown in panel (c).
}
\label{fig6}
\end{figure}

The spectra undergo significant changes as the field is applied, see
Fig.~\ref{fig6}. Local spectral functions (top panel) reveal that
the spin resonances are resilient to small fields and that their
position remains roughly constant as $B$ is increased. They are washed
away at higher fields when the AFM order itself becomes strongly
suppressed. This is in line with the interpretation of the spin
resonance as a manifestation of the staggered magnetization. The width
of the resonances is, however, strongly field dependent, reaching a
maximum for values of order $B_m$ where the staggered
magnetization peaks. When $B$ increases further, the resonance is
suppressed at the same time as the staggered magnetization tends
toward 0, as expected. The behavior of spectral functions near the
band edges is equally interesting. In particular, we note the
reemergence of the structure characteristic for the weak-coupling case
with square-root and inverse-square-root singularities. 

The field-dependence can be better understood through the
momentum-resolved spectral functions, see panel (b) in
Fig.~\ref{fig6}. In weak field, the main effect is the ``doubling'' of
the quasiparticle branches (four to eight). This results from the
breaking of the symmetry relation
$G_{A\sigma}(\omega)=G_{B\bar{\sigma}}(\omega)$ which guarantees the
degeneracy of the branches in the absence of the external field.
Physically, this means that in the presence of the field the $c$ band
electrons propagate slightly differently if their spin has a
transverse component which is aligned or antialigned with the uniform
component of the magnetization. This difference becomes more
pronounced at larger fields, and the splitting grows larger (see the
case of $B/D=0.12$). We also note that the higher-energy branches
(label H in the plot) always have much shorter quasiparticle lifetime
than the lower-energy ones (label L in the plot) because of the
relaxation mechanism via transverse spin component reorientation,
taking the quasiparticles from the upper to the lower branch. At high
fields the H branches become so diffuse that they can hardly be
distinguished. This evolution can also be followed in the
constant-momentum section of the momentum-resolved spectrum shown in
Fig.~\ref{fig6}(c). 

A further effect of the field is the emergence of the curvature in the
L branches, see label M in Fig.~\ref{fig6}(b). This new feature
directly explains the resurgence of the (inverse)-square-root
singularities at the gap edges, since the direct gap moves from the
non-interacting band edges at $\epsilon=\pm D$, where the DOS goes to
zero, $\rho(\pm D)=0$, to inner regions, gradually shifting to the
center of the band at $\epsilon=0$ as $B$ increases.

In Sec.~\ref{disc} we will see that most of these features can be
explained in the hybridization picture with longitudinal uniform and
transverse staggered magnetization.

\subsection{Universality and robustness}
\label{univ}

It has been pointed out that in the DMFT the most important characteristic
of the non-interacting density of states (DOS) of the lattice is its
effective bandwidth, defined through the second moment of the DOS,
\begin{equation}
D_\mathrm{eff} = \int\, \epsilon^2 \rho_0(\epsilon) \mathrm{d}\epsilon,
\end{equation}
which sets the scale of the kinetic energy. It is equal to $D$ for the
Bethe lattice and 2D cubic (square) lattice, $0.816D$ for 3D cubic
lattice, and $1.41D$ for the hypercubic lattice. Indeed, it has been
found that the Mott metal-insulator-transition at $T=0$ in the
paramagnetic phase of the Hubbard model occurs at roughly the same
value of the rescaled electron-electron repulsion parameter
$U/D_\mathrm{eff}$, which reflects the nature of the transition:
competition between the delocalizing effect of the kinetic energy and
the localizing effect of the electron-electron repulsion.

For the AFM-KI phase transition in the KLM, we also find that the
critical coupling is given by essentially the same ratio of
$J_c/D_\mathrm{eff}$ (we obtain $J_c/D_\mathrm{eff} \approx 0.56$ for
the Bethe lattice, 2D and 3D cubic, and $J_c/D_\mathrm{eff} \approx
0.54$ for the hypercubic lattice). This can again be rationalized in
terms of a competition between kinetic and exchange terms: kinetic
terms promote delocalization of $c$ electrons, while the exchange
terms enhance their localization by generating localized Kondo singlet
states. This essentially agrees with Doniach's picture of competing
RKKY and Kondo ground states.

\begin{figure} \centering
\includegraphics[clip,width=0.48\textwidth]{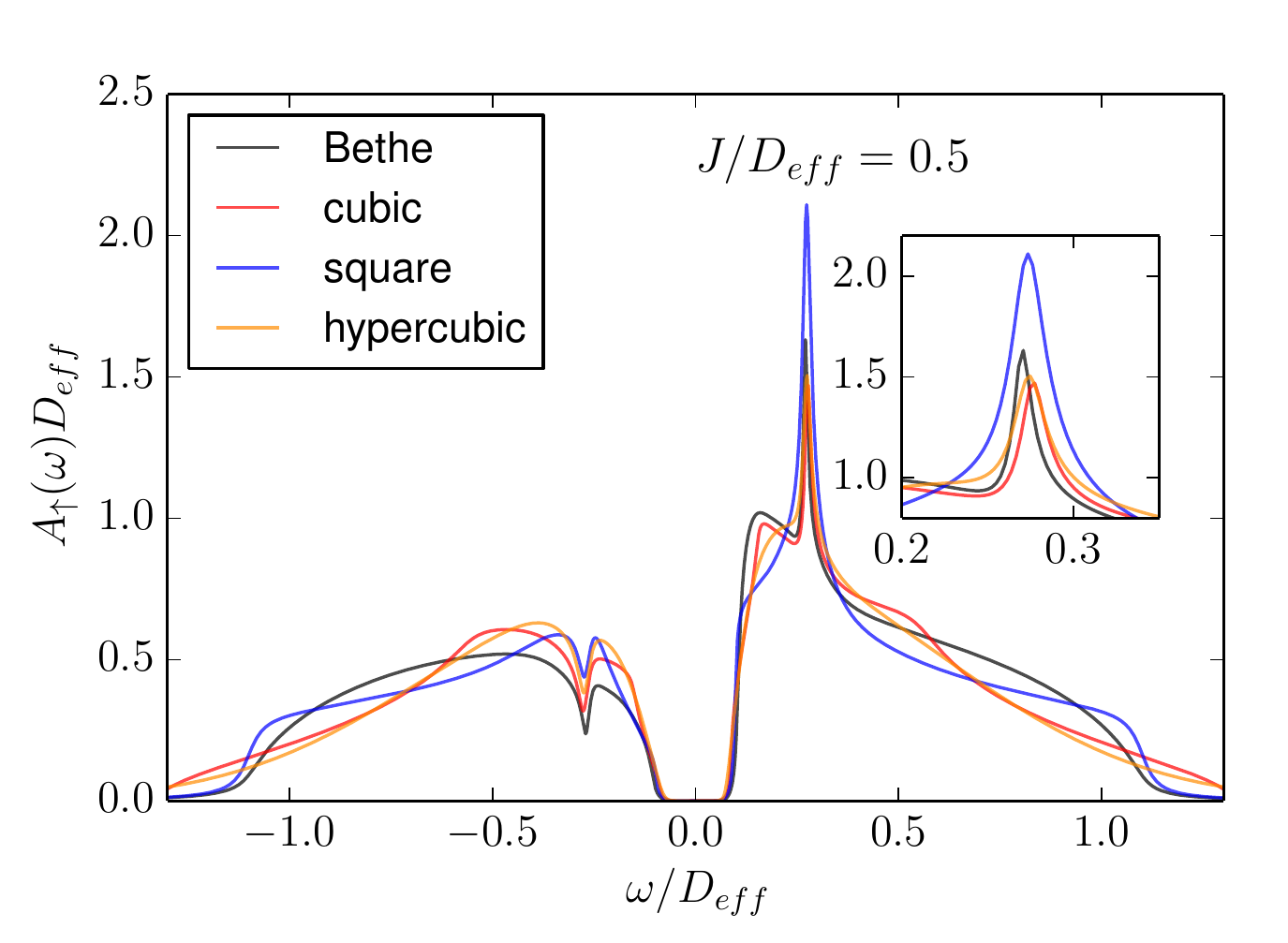}
\caption{(Color online) Spectral functions of the Kondo lattice model
on four different lattices: Bethe lattice, 2D cubic (square lattice),
3D cubic, and infinite-D cubic (Gaussian DOS) lattices. Note that the
figure axes are scaled in terms of the effective bandwidth
$D_\mathrm{eff}$ and that the same rescaled parameter
$J/D_\mathrm{eff}$ has been used in all four DMFT calculations.}
\label{lattice}
\end{figure}

The scaling in terms of $D_\mathrm{eff}$ is valid even more
generally. The comparison of spectral functions computed for different
lattice types, Fig.~\ref{lattice}, shows that despite significant
differences in details, the main features of appropriately rescaled
spectral functions are common to all cases: a) they have essentially
the same quasiparticle gap $\Delta$, b) they exhibit a spin resonance
structure, and c) the spin resonance appears at roughly the same
frequency $\omega_{sr}$ and has comparable spectral weight (with the
exception of square lattice which has a van Hove singularity at
$\epsilon=0$ that enhances the spin resonance peak).

\begin{figure} \centering
\includegraphics[clip,width=0.48\textwidth]{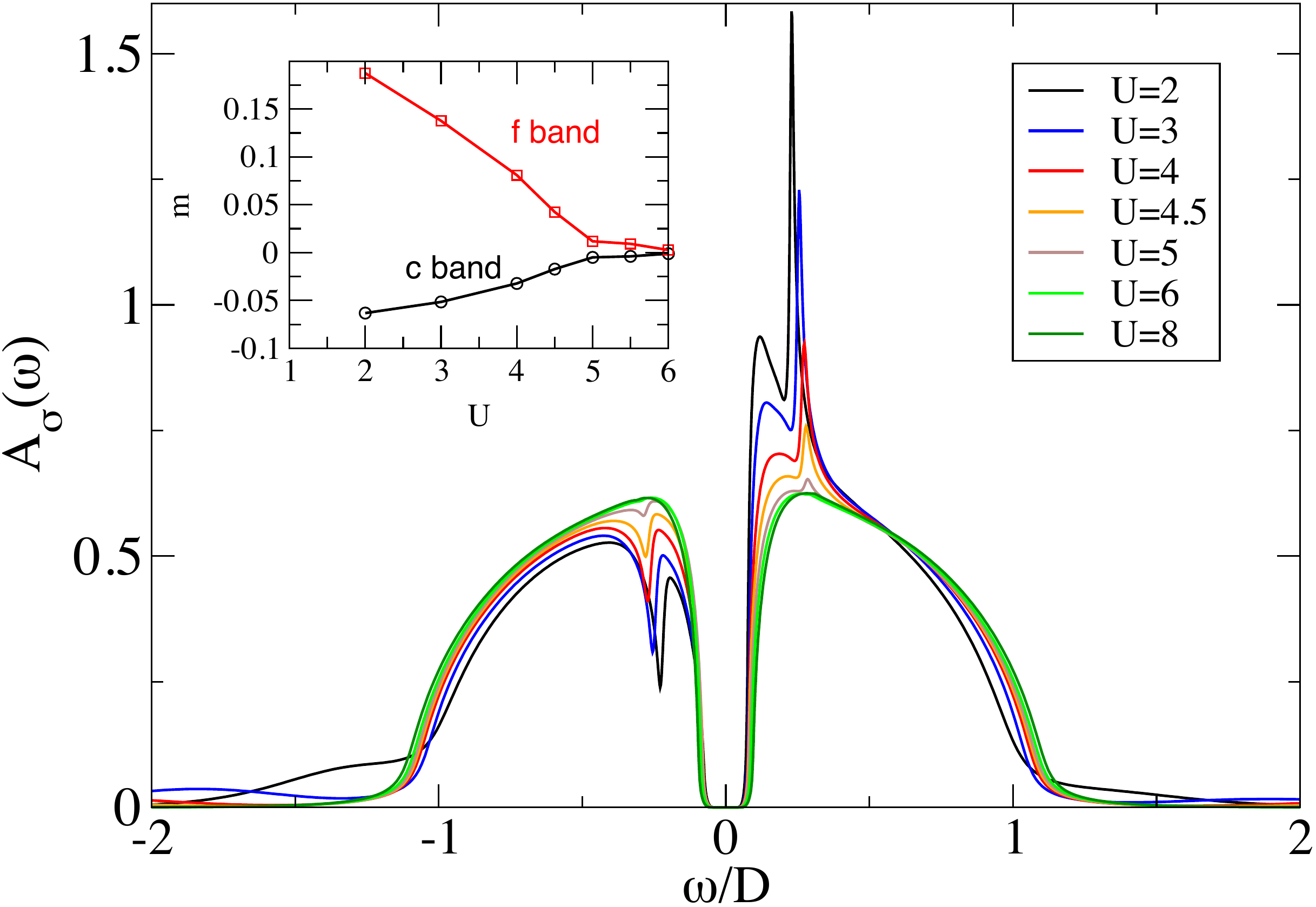}
\caption{(Color online) In periodic Anderson model (PAM), if $J
\propto t^2/U$ is kept constant while $t$ and $U$ increase, the AFM
order is suppressed, the spin resonance disappears, yet the gap
remains constant.}
\label{pam}
\end{figure}

The spin resonance is also present in closely related models which
have itinerant AFM order: high-spin Kondo lattice model
(explicitly tested for $S=1$ KLM, also in presence of magnetic
anisotropy term $DS_z^2$, for both axial $D<0$ and planar $D>0$
anisotropy) and the periodic Anderson model (PAM) with parameters
chosen so that the model is particle-hole symmetric and in the Kondo
limit (large $U$ and $\epsilon+U/2=0$). In PAM, if hybridization
$t$ and $f$-level charge repulsion $U$ are increased while keeping
the effective Kondo coupling $J \propto t^2/U$ constant, the
hybridization is increased in comparison with the exchange energy. The
result is that the staggered magnetization decreases and the spin
resonance gradually disappears, yet the spectral gap remains roughly
constant, see Fig.~\ref{pam}. Interestingly, as $t$ increases at
constant $J$, the charge fluctuations on the $f$ level actually
decrease due to increasing $U$.

\section{Discussion}
\label{disc}

The DMFT results indicate that at half filling the hybridization
picture is an essentially correct description of the antiferromagnetic
phase of the Kondo lattice model and that the topology of the
quasiparticle bands remains the same (large Fermi surface) for all
values of $J$. At the same time, our numerical results indicate that
at the quantitative level there are interesting details that have
experimentally observable consequences, such as the presence of
enhanced and suppressed density of states in the centre of the band at
the avoided crossing points of the quasiparticle bands (visible in
ARPES) and the non-trivial structure of the optical conductivity (see,
in particular, the comparison in Fig.~\ref{afmpm}). The hybridization
picture does not include any inelastic-scattering processes, since it
is essentially a non-interacting theory. Even at $T=0$ it therefore
does not properly capture effects away from the Fermi level, but
nevertheless it is a good starting point.

Let us first analyze the equation for the quasiparticle bands,
\begin{equation}
\Re[\zeta_{A\sigma}(\omega)\zeta_{B\sigma}(\omega)-\epsilon^2]=0,
\end{equation}
focusing on the region close to the spin resonance at $\epsilon=0$. We
are then actually solving
\begin{equation}
\Re[(\omega-\Sigma_\uparrow(\omega))(\omega-\Sigma_\downarrow(\omega))]=0.
\end{equation}
Neglecting the imaginary parts of $\Sigma$, this reduces to
\begin{equation}
(\omega-\Re\Sigma_\uparrow(\omega))(\omega-\Re\Sigma_\downarrow(\omega))]=0,
\end{equation}
and it follows that the solutions are given by
\begin{equation}
\omega = \Re\Sigma_\sigma(\omega).
\end{equation}
It turns out that in the range of $J$ where the spin resonance is the
most pronounced, this equation has solutions at $\omega \approx \pm
\omega_\mathrm{sr}$ {\it for both spin directions}. In other words,
one has
\begin{equation}
\Re\Sigma_\uparrow(\omega_\mathrm{sr}) \approx \Re\Sigma_\downarrow(\omega_\mathrm{sr})
\approx \omega_\mathrm{sr}.
\end{equation}
This condition has been interpreted in Ref.~\onlinecite{hoshino2010} in terms of
the molecular fields $h$ and $H$ as the relation $h=-H$ and named the
``quasilocal compensation''. We find, however, that this
``compensation'' is not generally valid.

\begin{figure} \centering
\includegraphics[clip,width=0.48\textwidth]{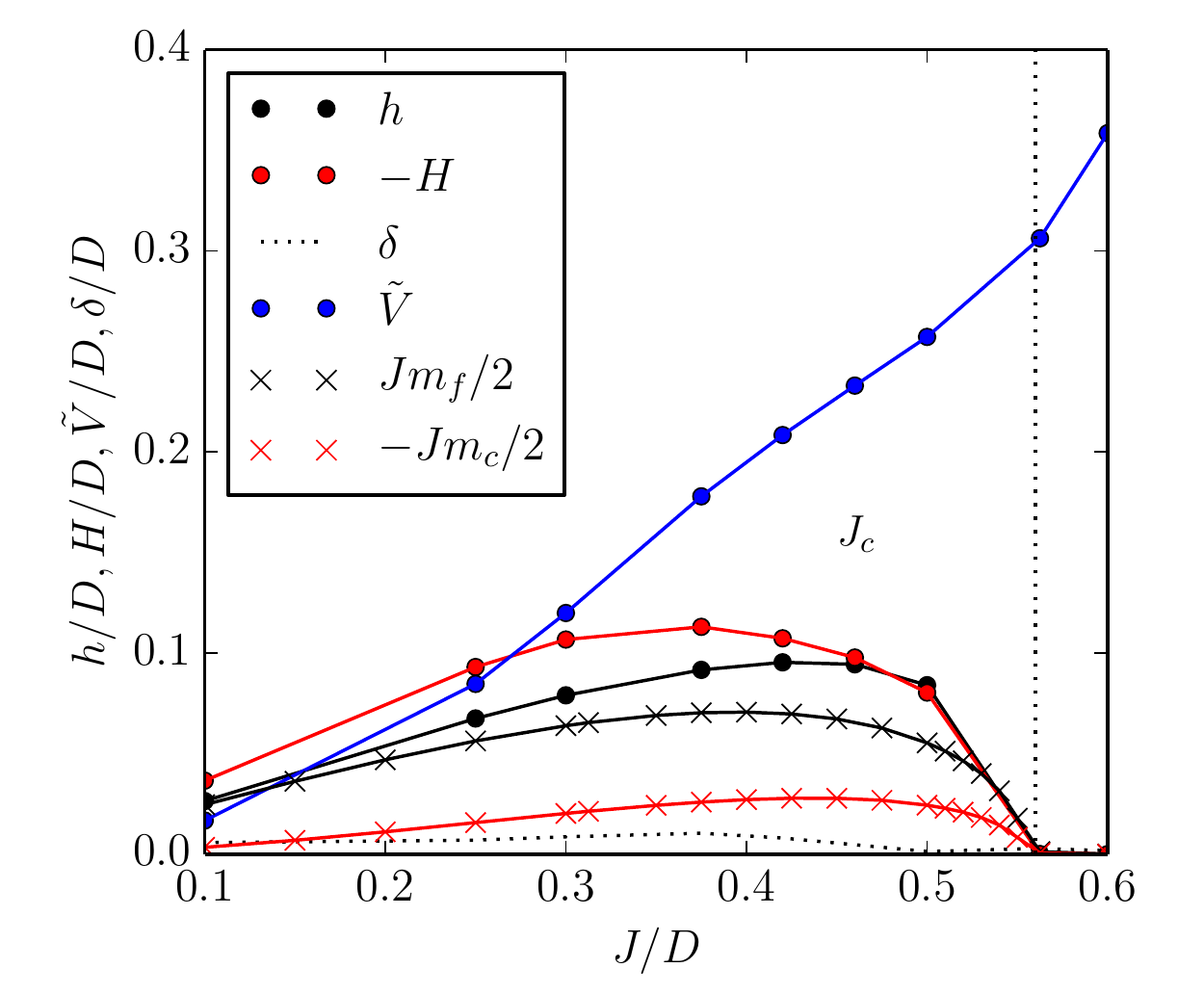}
\caption{(Color online) Kondo antiferromagnet in the hybridization
picture with spin polarization: parameters for the single-pole Ansatz
for the self-energy $\Sigma_\sigma(\omega)$ as a function of the Kondo
coupling $J$.}
\label{fig8}
\end{figure}

We have systematically extracted the parameters from the calculated
self-energy functions using the following hybridization-picture
Ansatz:
\begin{equation}
\Sigma_\uparrow(\omega) = h + \frac{\tilde{V}^2}{z-H+i\delta}.
\end{equation}
A small imaginary part $\delta$ has been added to account for the
finite width of the peak in $\Im \Sigma$. For stability, the parameter
extraction has been performed simultaneously on real and imaginary
parts of the function. The results are shown in Fig.~\ref{fig8}. The
plot reveals that the curves $h(J)$ and $-H(J)$ have a similar
non-monotonic behavior with a maximum value in the cross-over region
between the weak- and strong-coupling antiferromagnet, but they
intersect at a single plot near $J/D=0.5$. The hybridization parameter
$\tilde{V}$ is continuous across the AFM-KI transition, as already
noted in Ref.~\onlinecite{hoshino2010}. The imaginary-part parameter $\delta$ is
small, but needs to be included for a good fit, even though it leads
to worse agreement with $\Im\Sigma(\omega)$, which, in particular,
should be strictly equal to zero inside the gap.

\begin{figure} \centering
\includegraphics[clip,width=0.48\textwidth]{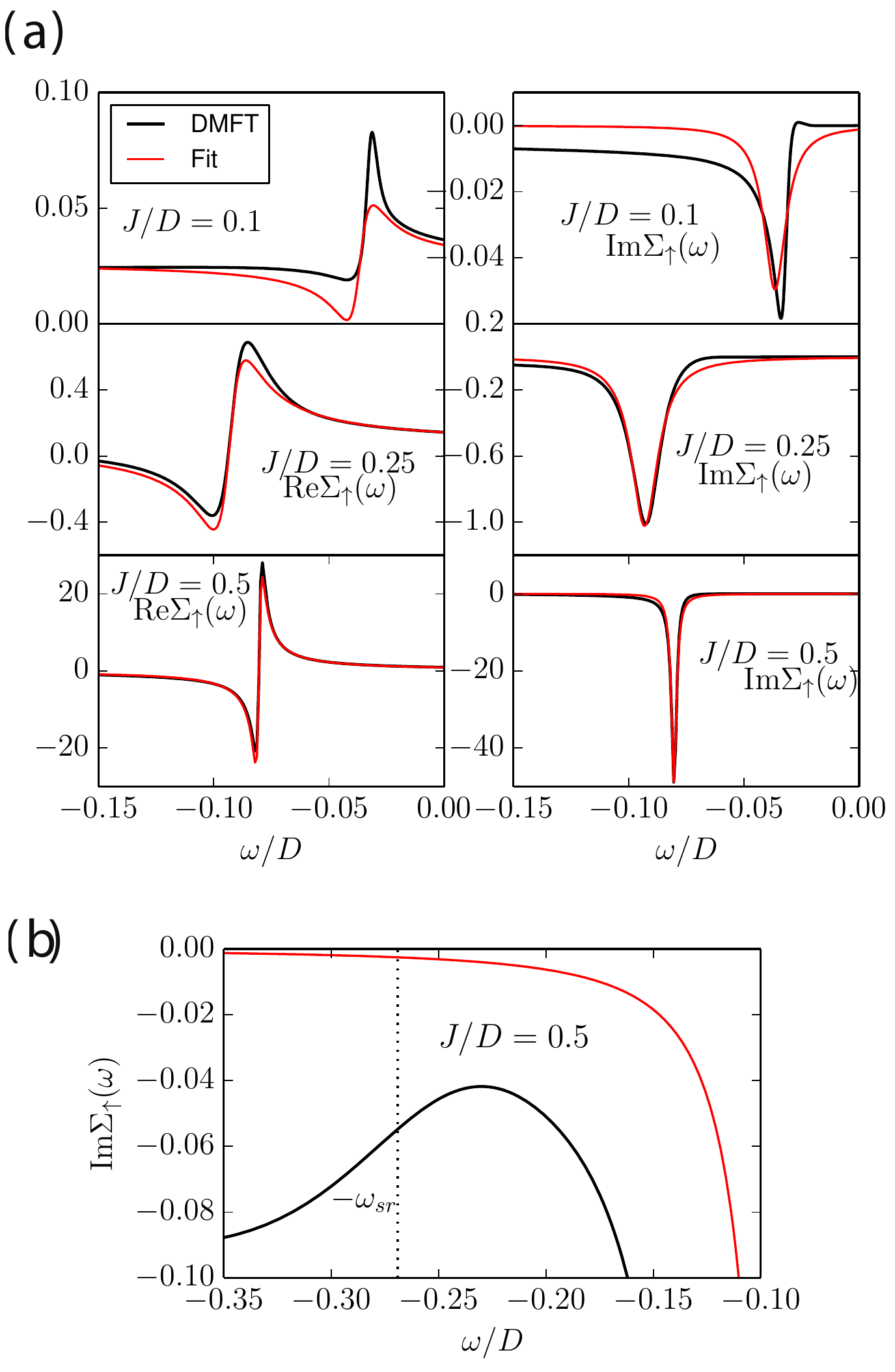}
\caption{(Color online) Self-energy functions and fits to the
hybridization-picture Ansatz with a single-pole.  The fit quality is
mediocre in the weak-coupling regime ($J/D=0.1$), but improves in the
strong-coupling regime ($J/D=0.5$). (b) $\Im\Sigma$ has some fine structure in
addition to the dominant poles.}
\label{fig9}
\end{figure}

In Fig.~\ref{fig9}(a) we plot the real and imaginary parts of the
self-energy together with the corresponding single-pole fit functions.
The agreement is better in the strong-coupling regime at $J/D=0.5$
where the pole is very strong and dominates the remaining structure in
the self-energy, visible in the close-up on $\Im\Sigma(\omega)$ shown
in panel (b) and labeled as B. Surprisingly, in the weak-coupling
regime at $J/D=0.1$, the agreement is much less satisfactory. The main
reason is that the pole is not much larger compared with the remaining
structure: region B is merged with the pole A, thus the peak is no
longer a simple Lorentzian and consequently $\Re\Sigma(\omega)$ 
is asymmetric.

\begin{figure} \centering
\includegraphics[clip,width=0.48\textwidth]{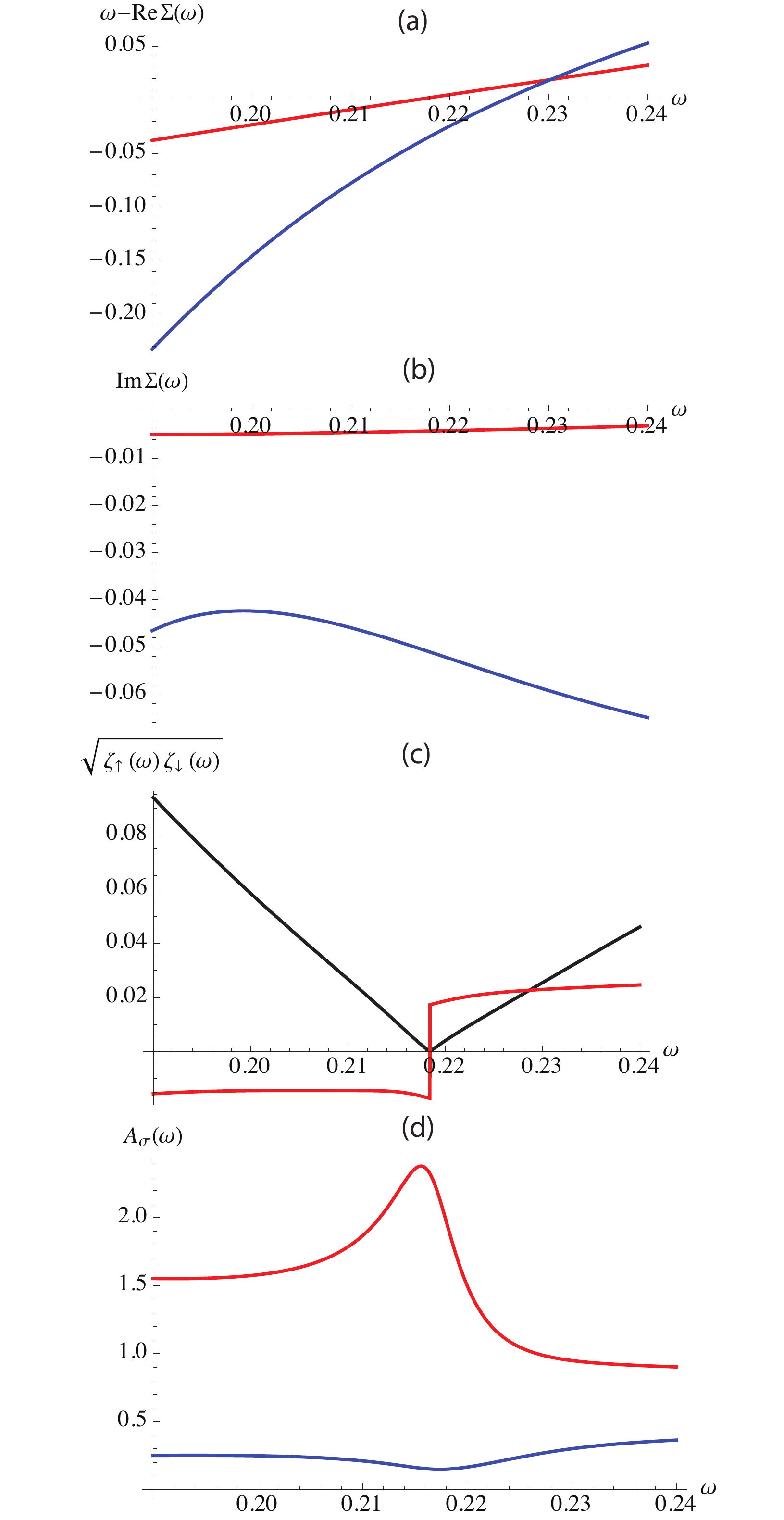}
\caption{(Color online) Analytical structure leading to the spin
resonance. a) Nearby solutions of $\omega=\Re\Sigma_\sigma(\omega)$
for $\sigma=\uparrow$ and $\sigma=\downarrow$. b)
$\Im\Sigma_\sigma(\omega)$ are nearly constant for $\omega \approx
\omega_\mathrm{sr}$. c) Argument of the non-interacting Green's
functions in the DMFT expression (Eq.~\eqref{eq22}) for local Green's
functions. d) Resulting local spectral functions featuring enhancement
or suppression at $\omega \approx \omega_\mathrm{sr}$.}
\label{X}
\end{figure}

We emphasize that the spin resonances are not located at the
frequencies of the poles in the self-energy, but at significantly
higher energies. We now study this in more detail by considering the
generic case at $J/D=0.4$ where $h$ and $-H$ differ slightly.
Functions
$Re\zeta_{A\sigma}(\omega)=\omega-\Re\Sigma_{A\sigma}(\omega)$
intersect the real axis at two different points, see Fig.~\ref{X}(a).
Somewhere between these two points, the function
$p=(\zeta_{A\sigma}(\omega) \zeta_{B\sigma}(\omega))^{1/2}$ goes
through a branch cut so that its imaginary part has a jump,
Fig.~\ref{X}(c). This discontinuity is canceled by that in
$G_0(p)-G_0(-p)$, resulting in a continuous spectral function, which
however has a peak, Fig.~\ref{X}(d).

\begin{figure} \centering
\includegraphics[clip,width=0.48\textwidth]{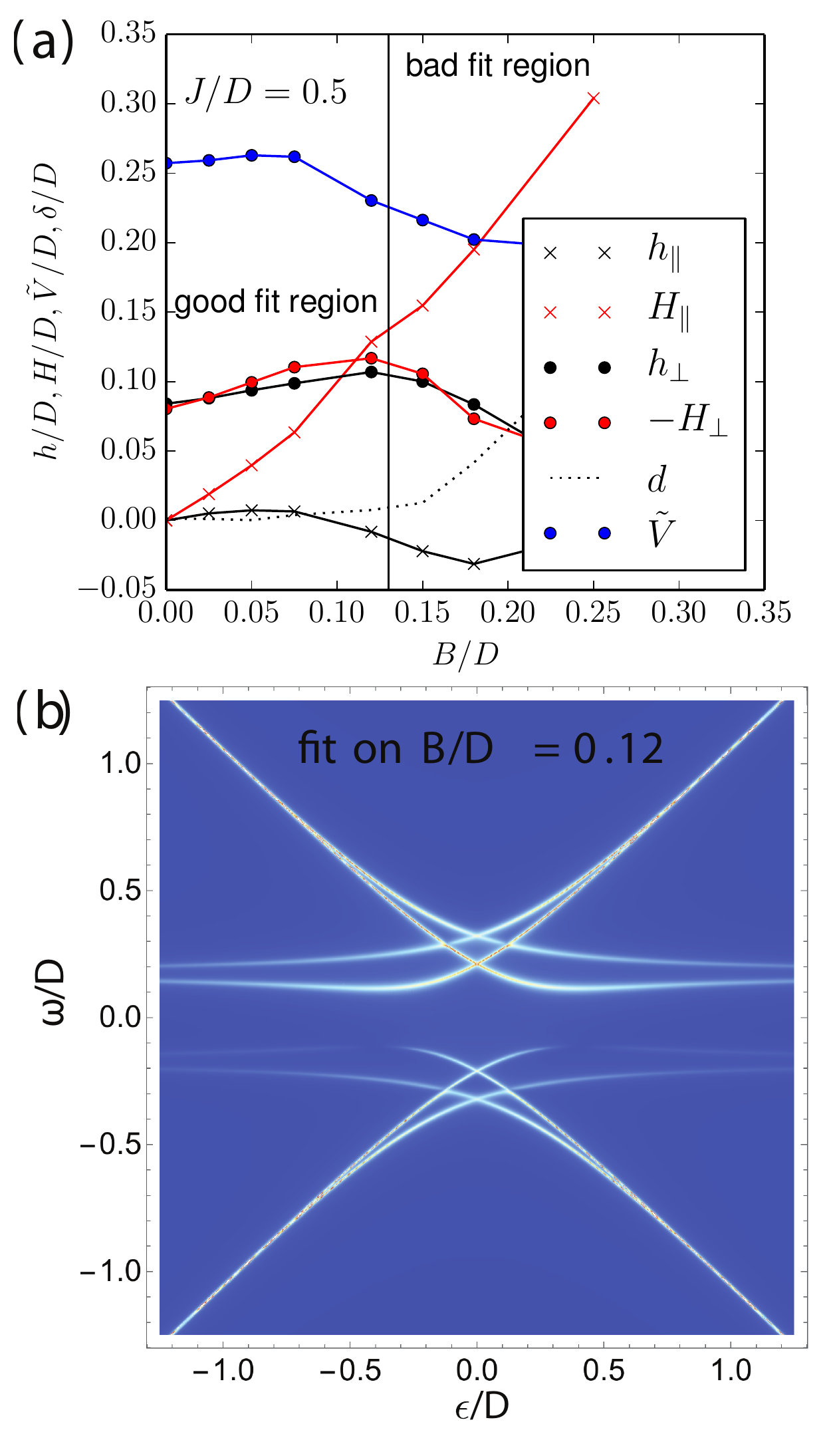}
\caption{(Color online) a) Parameters for the hybridization-picture
Ansatz for the Kondo antiferromagnet in external magnetic field.
b) Momentum-resolved spectral function at finite magnetic field.
}
\label{figlast}
\end{figure}

We now consider the case of finite magnetic field where the
momentum-resolved spectral functions show complex structure with
quasiparticle branch doubling in number. The simplest attempt to
rationalize this behavior is to incorporate the additional uniform
magnetic field in the hybridization picture. The self-energy function
now has a $2 \times 2$ matrix structure:
\begin{equation}
\boldsymbol{\Sigma}(\omega) = 
\begin{pmatrix}
h_\perp & h_\| \\
h_\| & - h_\perp
\end{pmatrix}
+
\tilde{V}^2 \left( z - \begin{pmatrix}
H_\perp & H_\| \\
H_\| & -H_\perp
\end{pmatrix}
+ i \delta \right)^{-1}.
\end{equation}
The extracted parameters are shown in Fig.~\ref{figlast}. The
staggered components $h_\perp$ and $H_\perp$ (which correspond to the
previously discussed $h$ and $H$ in the absence of the field) have
only weak $B$ dependence: at first they slightly grow (in absolute
value), similar to the staggered magnetization components $m_{c\perp}$
and $m_{f\perp}$, then decrease. The homogeneous longitudinal
molecular field components also mimick the corresponding magnetization
components: $H_\|$ rapidly grows with $B$ and eventually becomes the
dominant molecular field, while $h_\|$ slightly increases and then
changes sign. The effective hybridization $\tilde{V}$ does not change
appreciably with field. We also note that the quality of the fit
worsens at high fields. This is expected since the systems
renormalizes toward a weakly-interacting spin-polarized limit where
the simple hybridization picture is not a good approximation (similar
to the case of small $J$ at $B=0$).

It is worth mentioning that
in the strong-coupling regime for large $J/D \sim 0.5$, $\tilde{V}$ is
much larger than $h \approx -H$, thus $\omega^* =
\sqrt{\tilde{V}^2+h^2} \sim \tilde{V}$. The effective hybridization
$\tilde{V}$ is not strongly affected by the field, therefore
$\omega^*$ remains approximately constant. This explains why the spin
resonance position is not much affected by the external magnetic
field, as seen in Fig.~\ref{fig6}.

\section{Conclusion}

We have performed a detailed study of the spectral properties of the
Kondo lattice model at half-filling where the system is an itinerant
antiferromagnetic insulator for $J<J_c$ and a paramagnetic Kondo
insulator for $J>J_c$. The dynamical mean-field theory calculations
have been performed with a quantum impurity solver with high spectral
resolution. We have uncovered fine structure (``spin resonances'')
inside the bands at frequencies given by the crossing point of the
quasiparticle branches in the centre of the non-interacting band
($\epsilon=0$). These features are due to the inelastic-scattering
processes which are not taken into account in the simplified
hybridization picture. They are directly related to the existence of
the AFM order: whenever the AFM order disappears, either due to
thermal phase transition, external magnetic field, or quantum phase
transition to the KI, the spin resonances also disappear.

Similiar spin resonances also exist in the superconducting phase of
the KLM \cite{PhysRevLett.110.146406} and they share some other common
properties, for instance their position also changes linearly with
$J$. These analogies are not too surprising since the Nambu formalism
used to describe superconductivity is very similar to the A/B
sublattice formalism used to describe N\'eel order on bipartite
lattices, thus the analytical structure of the DMFT self-consistency
equations is analogous. Following this analogy, the resonances in the
superconducting case can be interpreted to be arising from
simultaneous presence of the $f$-electron itinerancy (heavy Fermi
liquid) and non-zero order parameter, and should thus appear 
generically in heavy-fermion $s$-wave superconductors. This is a
further indication that the superconducting state emerges out of the
large Fermi surface heavy fermion state.

The most direct way to experimentally observe such sharp spectral
structure is tunneling spectroscopy which gives access to local
spectral function where the ``spin resonance'' is the most pronounced
(more than in the momentum-resolved spectral functions measurable by
ARPES). The feature to look for is the apparition of an antisymmetric
resonance at finite frequencies as the system becomes
antiferromagnetic, with an intensity directly related to the order
parameter. Another possibility is optical spectroscopy where fine
details of peak shapes can also be easily measured.

The effect of non-local correlations on the observed fine structure
remains an open question and will be a part of future research. It
could be addressed, for instance, in cellular DMFT studies
\cite{Tanaskovic:2011jk}.

\begin{acknowledgments}
R. \v{Z}. and \v{Z}. O. acknowledge the support of the Slovenian
Research Agency (ARRS) under Program No. P1-0044 and T. P. the support
through the German Science Foundation through project PR 298/13-1. We
acknowledge discussions with O. Bodensiek who participated in the
early stages of this work.
\end{acknowledgments}

\appendix

\begin{widetext}

\section{DMFT(NRG) approach in the absence of spin symmetries}
\label{appA}

The inverse Green's function has a block $2 \times 2$ matrix
structure, each block being itself a $2 \times 2$ matrix in the spin
space:
\begin{equation}
G_k^{-1}(z) = 
\begin{pmatrix}
z + \mu  - \tau^3 (h+h_s) -\tau^1 h_t - \Sigma_A & -\epsilon_k \\
-\epsilon_k & z + \mu - \sigma^3 (h-h_s) - \tau^1 h_t -\Sigma_B
\end{pmatrix}.
\end{equation}
Here $\tau^i$ are Pauli matrices, $h_s$ is the staggered field, $h$
is the longitudinal homogeneous field and $h_t$ is the homogenous
transverse field. As in the spin diagonal case, we introduce $\zeta_A$
and $\zeta_B$,
\begin{equation}
G_k^{-1}(z) =
\begin{pmatrix}
\zeta_A & -\epsilon_k \\
-\epsilon_k & \zeta_B
\end{pmatrix}.
\end{equation}
We assume $\zeta_A$ and $\zeta_B$ to be invertible, and perform a
blockwise invertion of the matrix:
\begin{equation}
 M=\begin{pmatrix}
 A & B \\
 C & D 
 \end{pmatrix}
 \quad
 \leftrightarrow
 M^{-1} =\begin{pmatrix}
 (A-BD^{-1}C)^{-1} & -A^{-1}B(D-CA^{-1}B)^{-1} \\
 -D^{-1}C(A-BD^{-1}C)^{-1} & (D-CA^{-1}B)^{-1}
 \end{pmatrix}
 \end{equation}
 thus
 \begin{equation}
 G_k(z) = 
 \begin{pmatrix}
 (\zeta_A-\epsilon_k^2 \zeta_B^{-1})^{-1} & {\ldots}  \\
 {\ldots}  & (\zeta_B-\epsilon_k^2 \zeta_A^{-1})^{-1}.
 \end{pmatrix}
 \end{equation}
 The out-of-diagonal elements are of no interest, because they are odd
 functions of $\epsilon_k$ and will drop out after the integration
 since $D(\epsilon)$ is assumed to be even (It is indeed even for all
 lattice types considered in this work).

 The local Green's function is
 \begin{equation}
 G(z)=\frac{1}{N}\sum_k G_k(z) = 
 \int d\epsilon D(\epsilon) 
 \begin{pmatrix}
 \zeta_B (\zeta_A \zeta_B-\epsilon^2)^{-1} & {\ldots}  \\
 {\ldots} & \zeta_A (\zeta_B \zeta_A-\epsilon^2)^{-1}
 \end{pmatrix}.
 \end{equation}
 Note that $\zeta_A$ and $\zeta_B$ in general do not commute.
 
 We consider each diagonal submatrix problem. We write
 \begin{equation}
 F^A = \zeta_A \zeta_B, \quad F^B = \zeta_B \zeta_A
 \end{equation}
 and
 \begin{equation}
 D^A = F^A-\epsilon^2, \quad D^B = F^B-\epsilon^2.
 \end{equation}
 
 We need to integrate each matrix component separately, but the pole
 structure is the same for all components.
 
 We write
 \begin{equation}
 F^A = \begin{pmatrix}
 F_{11} & F_{12} \\
 F_{21} & F_{22}
 \end{pmatrix}
 \end{equation}
 and
 \begin{equation}
 D^A = \begin{pmatrix} F_{11}-\epsilon^2 & F_{12} \\ 
 F_{21} & F_{22}-\epsilon^2 \end{pmatrix}.
 \end{equation}
 Then
 \begin{equation}
 [D^A]^{-1} = 
 \frac{1}{(F_{11}-\epsilon^2)(F_{22}-\epsilon^2)-F_{12}F_{21}} 
 \begin{pmatrix}
 F_{22}-\epsilon^2 & -F_{12} \\
 -F_{21} & F_{11}-\epsilon^2
 \end{pmatrix}.
 \end{equation}
 
 We expand the fraction:
 \begin{equation}
 \frac{1}{(F_{11}-\epsilon^2)(F_{22}-\epsilon^2)-F_{12}F_{21}} 
 =c^A
 \left[ 
 \frac{1/\epsilon_1}{\epsilon-\epsilon_1}
 +
 \frac{-1/\epsilon_1}{\epsilon+\epsilon_1}
 +
 \frac{-1/\epsilon_2}{\epsilon-\epsilon_2}
 +
 \frac{1/\epsilon_2}{\epsilon+\epsilon_2}
 \right],
 \end{equation}
 where
 \begin{equation}
 \epsilon_{1,2} = \frac{1}{\sqrt{2}} \left(
 F_{11}+F_{22} \pm \sqrt{F_{11}^2 + F_{22}^2 + 4 F_{12} F_{21}
 -2F_{11}F_{22} } \right)^{1/2}
 \end{equation}
 and
 \begin{equation}
 c^A = \frac{1/2}{(\epsilon_1-\epsilon_2)(\epsilon_1+\epsilon_2)}.
 \end{equation}
 We use the relation
 \begin{equation}
 \int \frac{D(\epsilon)\epsilon^2d\epsilon}{z-\epsilon}
 =-z+z^2 \int \frac{D(\epsilon)d\epsilon}{z-\epsilon}
 =-z+z^2 G^0(z)
 \end{equation}
 where $G^0(z)$ is the non-interacting local Green's function for the
 chosen lattice problem. 
 Thus, for example
 \begin{equation}
 \int \frac{D(\epsilon)d\epsilon}{\epsilon-\epsilon_1} =
 -G^0(\epsilon_1),
 \end{equation}
 and
 \begin{equation}
 \int \frac{\epsilon^2 D(\epsilon)d\epsilon}{\epsilon-\epsilon_1} =
 \epsilon_1-\epsilon_1^2 G^0(\epsilon_1).
 \end{equation}
 
 Then
 \begin{equation}
 G(z) = \int d\epsilon D(\epsilon) 
 \begin{pmatrix}
 \zeta_B [D^A]^{-1} & {\ldots}  \\
 {\ldots}  &\zeta_A  [D^B]^{-1}
 \end{pmatrix}
 =
 \begin{pmatrix}
 \zeta_B J^A & 0  \\
 0 & \zeta_A J^B
 \end{pmatrix},
 \end{equation}
 where $J^{A/B}$ are the integrals over $\epsilon$. Since 
 $\zeta_{A/B}$ depend only on $z$, not $\epsilon$, they may be
 factored out and taken into account after the integration.
 
 For $J^A$ we find
 \begin{equation}
 \begin{split}
 J^A &= c^A 
 \begin{pmatrix}
 F_{22} & -F_{12} \\
 -F_{21} & F_{11}
 \end{pmatrix}
 \left[ 
 -(1/\epsilon_1) G^0(\epsilon_1)
 +(1/\epsilon_1) G^0(-\epsilon_1)
 +(1/\epsilon_2) G^0(\epsilon_2)
 -(1/\epsilon_2) G^0(-\epsilon_2)
 \right] \\
 & -c^A
 \begin{pmatrix}
 1 & 0 \\
 0 & 1
 \end{pmatrix}
 \left\{
 +(1/\epsilon_1)[\epsilon_1-\epsilon_1^2 G^0(\epsilon_1)]
 -(1/\epsilon_1)[-\epsilon_1-\epsilon_1^2 G^0(-\epsilon_1)]
 -(1/\epsilon_2)[\epsilon_2-\epsilon_2^2 G^0(\epsilon_2)]
 +(1/\epsilon_2)[-\epsilon_2-\epsilon_2^2 G^0(-\epsilon_2)]
 \right\}.
 \end{split}
 \end{equation}

 For each $A/B$ subproblem, the hybridization function is then the
 standard one:
 \begin{equation}
 \Delta_i(z) = \Im\left[ G_i^{-1}(z) + \Sigma_i(z) \right],
 \end{equation}
 with $i=A/B$, and $\Delta_i$, $G_i$ and $\Sigma_i$ are all 2x2
 matrices.

\end{widetext}

\bibliography{refs}

\begin{thebibliography}{38}
\expandafter\ifx\csname natexlab\endcsname\relax\def\natexlab#1{#1}\fi
\expandafter\ifx\csname bibnamefont\endcsname\relax
  \def\bibnamefont#1{#1}\fi
\expandafter\ifx\csname bibfnamefont\endcsname\relax
  \def\bibfnamefont#1{#1}\fi
\expandafter\ifx\csname citenamefont\endcsname\relax
  \def\citenamefont#1{#1}\fi
\expandafter\ifx\csname url\endcsname\relax
  \def\url#1{\texttt{#1}}\fi
\expandafter\ifx\csname urlprefix\endcsname\relax\def\urlprefix{URL }\fi
\providecommand{\bibinfo}[2]{#2}
\providecommand{\eprint}[2][]{\url{#2}}

\bibitem[{\citenamefont{Hewson}(1997)}]{hewson1997kondo}
\bibinfo{author}{\bibfnamefont{A.~C.} \bibnamefont{Hewson}},
  \emph{\bibinfo{title}{The Kondo problem to heavy fermions}},
  \bibinfo{number}{2} (\bibinfo{publisher}{Cambridge university press},
  \bibinfo{year}{1997}).

\bibitem[{\citenamefont{Stewart}(1984)}]{RevModPhys.56.755}
\bibinfo{author}{\bibfnamefont{G.~R.} \bibnamefont{Stewart}},
  \bibinfo{journal}{Rev. Mod. Phys.} \textbf{\bibinfo{volume}{56}},
  \bibinfo{pages}{755} (\bibinfo{year}{1984}),
  \urlprefix\url{http://link.aps.org/doi/10.1103/RevModPhys.56.755}.

\bibitem[{\citenamefont{Jaime et~al.}(2000)\citenamefont{Jaime, Movshovich,
  Stewart, Beyermann, Berisso, Hundley, Canfield, and
  Sarrao}}]{jaime2000closing}
\bibinfo{author}{\bibfnamefont{M.}~\bibnamefont{Jaime}},
  \bibinfo{author}{\bibfnamefont{R.}~\bibnamefont{Movshovich}},
  \bibinfo{author}{\bibfnamefont{G.~R.} \bibnamefont{Stewart}},
  \bibinfo{author}{\bibfnamefont{W.~P.} \bibnamefont{Beyermann}},
  \bibinfo{author}{\bibfnamefont{M.~G.} \bibnamefont{Berisso}},
  \bibinfo{author}{\bibfnamefont{M.~F.} \bibnamefont{Hundley}},
  \bibinfo{author}{\bibfnamefont{P.~C.} \bibnamefont{Canfield}},
  \bibnamefont{and} \bibinfo{author}{\bibfnamefont{J.~L.}
  \bibnamefont{Sarrao}}, \bibinfo{journal}{Nature}
  \textbf{\bibinfo{volume}{405}}, \bibinfo{pages}{160} (\bibinfo{year}{2000}).

\bibitem[{\citenamefont{Sugiyama et~al.}(1988)\citenamefont{Sugiyama, Iga,
  Kasaya, Kasuya, and Date}}]{sugiyama1988field}
\bibinfo{author}{\bibfnamefont{K.}~\bibnamefont{Sugiyama}},
  \bibinfo{author}{\bibfnamefont{F.}~\bibnamefont{Iga}},
  \bibinfo{author}{\bibfnamefont{M.}~\bibnamefont{Kasaya}},
  \bibinfo{author}{\bibfnamefont{T.}~\bibnamefont{Kasuya}}, \bibnamefont{and}
  \bibinfo{author}{\bibfnamefont{M.}~\bibnamefont{Date}},
  \bibinfo{journal}{Journal of the Physical Society of Japan}
  \textbf{\bibinfo{volume}{57}}, \bibinfo{pages}{3946} (\bibinfo{year}{1988}).

\bibitem[{\citenamefont{Mason et~al.}(1992)\citenamefont{Mason, Aeppli,
  Ramirez, Clausen, Broholm, St\"ucheli, Bucher, and Palstra}}]{mason1992}
\bibinfo{author}{\bibfnamefont{T.~E.} \bibnamefont{Mason}},
  \bibinfo{author}{\bibfnamefont{G.}~\bibnamefont{Aeppli}},
  \bibinfo{author}{\bibfnamefont{A.~P.} \bibnamefont{Ramirez}},
  \bibinfo{author}{\bibfnamefont{K.~N.} \bibnamefont{Clausen}},
  \bibinfo{author}{\bibfnamefont{C.}~\bibnamefont{Broholm}},
  \bibinfo{author}{\bibfnamefont{N.}~\bibnamefont{St\"ucheli}},
  \bibinfo{author}{\bibfnamefont{E.}~\bibnamefont{Bucher}}, \bibnamefont{and}
  \bibinfo{author}{\bibfnamefont{T.~T.~M.} \bibnamefont{Palstra}},
  \bibinfo{journal}{Phys. Rev. Lett.} \textbf{\bibinfo{volume}{69}},
  \bibinfo{pages}{490} (\bibinfo{year}{1992}).

\bibitem[{\citenamefont{Bat’ková et~al.}(2006)\citenamefont{Bat’ková,
  Bat’ko, Konovalova, Shitsevalova, and Paderno}}]{batkova2006}
\bibinfo{author}{\bibfnamefont{M.}~\bibnamefont{Bat’ková}},
  \bibinfo{author}{\bibfnamefont{I.}~\bibnamefont{Bat’ko}},
  \bibinfo{author}{\bibfnamefont{E.}~\bibnamefont{Konovalova}},
  \bibinfo{author}{\bibfnamefont{N.}~\bibnamefont{Shitsevalova}},
  \bibnamefont{and} \bibinfo{author}{\bibfnamefont{Y.}~\bibnamefont{Paderno}},
  \bibinfo{journal}{Physica B: Condensed Matter}
  \textbf{\bibinfo{volume}{378–380}}, \bibinfo{pages}{618 }
  (\bibinfo{year}{2006}).

\bibitem[{\citenamefont{Knafo et~al.}(2010)\citenamefont{Knafo, Aoki,
  Vignolles, Vignolle, Klein, Jaudet, Villaume, Proust, and
  Flouquet}}]{PhysRevB.81.094403}
\bibinfo{author}{\bibfnamefont{W.}~\bibnamefont{Knafo}},
  \bibinfo{author}{\bibfnamefont{D.}~\bibnamefont{Aoki}},
  \bibinfo{author}{\bibfnamefont{D.}~\bibnamefont{Vignolles}},
  \bibinfo{author}{\bibfnamefont{B.}~\bibnamefont{Vignolle}},
  \bibinfo{author}{\bibfnamefont{Y.}~\bibnamefont{Klein}},
  \bibinfo{author}{\bibfnamefont{C.}~\bibnamefont{Jaudet}},
  \bibinfo{author}{\bibfnamefont{A.}~\bibnamefont{Villaume}},
  \bibinfo{author}{\bibfnamefont{C.}~\bibnamefont{Proust}}, \bibnamefont{and}
  \bibinfo{author}{\bibfnamefont{J.}~\bibnamefont{Flouquet}},
  \bibinfo{journal}{Phys. Rev. B} \textbf{\bibinfo{volume}{81}},
  \bibinfo{pages}{094403} (\bibinfo{year}{2010}),
  \urlprefix\url{http://link.aps.org/doi/10.1103/PhysRevB.81.094403}.

\bibitem[{\citenamefont{Degiorgi}(1999)}]{degiorgi1996}
\bibinfo{author}{\bibfnamefont{L.}~\bibnamefont{Degiorgi}},
  \bibinfo{journal}{Reviews of Modern Physics} \textbf{\bibinfo{volume}{71}},
  \bibinfo{pages}{687} (\bibinfo{year}{1999}).

\bibitem[{\citenamefont{Wilson}(1975)}]{RevModPhys.47.773}
\bibinfo{author}{\bibfnamefont{K.~G.} \bibnamefont{Wilson}},
  \bibinfo{journal}{Rev. Mod. Phys.} \textbf{\bibinfo{volume}{47}},
  \bibinfo{pages}{773} (\bibinfo{year}{1975}),
  \urlprefix\url{http://link.aps.org/doi/10.1103/RevModPhys.47.773}.

\bibitem[{\citenamefont{Doniach}(1977)}]{doniach1977kondo}
\bibinfo{author}{\bibfnamefont{S.}~\bibnamefont{Doniach}},
  \bibinfo{journal}{Physica B+ C} \textbf{\bibinfo{volume}{91}},
  \bibinfo{pages}{231} (\bibinfo{year}{1977}).

\bibitem[{\citenamefont{Hoshino et~al.}(2010)\citenamefont{Hoshino, Otsuki, and
  Kuramoto}}]{hoshino2010}
\bibinfo{author}{\bibfnamefont{S.}~\bibnamefont{Hoshino}},
  \bibinfo{author}{\bibfnamefont{J.}~\bibnamefont{Otsuki}}, \bibnamefont{and}
  \bibinfo{author}{\bibfnamefont{Y.}~\bibnamefont{Kuramoto}},
  \bibinfo{journal}{Phys. Rev. B} \textbf{\bibinfo{volume}{81}},
  \bibinfo{pages}{113108} (\bibinfo{year}{2010}).

\bibitem[{\citenamefont{{Coleman}}(2006)}]{2006cond.mat.12006C}
\bibinfo{author}{\bibfnamefont{P.}~\bibnamefont{{Coleman}}},
  \bibinfo{journal}{eprint arXiv:cond-mat/0612006}  (\bibinfo{year}{2006}),
  \eprint{cond-mat/0612006}.

\bibitem[{\citenamefont{Georges
  et~al.}(1996{\natexlab{a}})\citenamefont{Georges, Kotliar, Krauth, and
  Rozenberg}}]{RevModPhys.68.13}
\bibinfo{author}{\bibfnamefont{A.}~\bibnamefont{Georges}},
  \bibinfo{author}{\bibfnamefont{G.}~\bibnamefont{Kotliar}},
  \bibinfo{author}{\bibfnamefont{W.}~\bibnamefont{Krauth}}, \bibnamefont{and}
  \bibinfo{author}{\bibfnamefont{M.~J.} \bibnamefont{Rozenberg}},
  \bibinfo{journal}{Rev. Mod. Phys.} \textbf{\bibinfo{volume}{68}},
  \bibinfo{pages}{13} (\bibinfo{year}{1996}{\natexlab{a}}),
  \urlprefix\url{http://link.aps.org/doi/10.1103/RevModPhys.68.13}.

\bibitem[{\citenamefont{Krishna-murthy
  et~al.}(1980)\citenamefont{Krishna-murthy, Wilkins, and
  Wilson}}]{PhysRevB.21.1003}
\bibinfo{author}{\bibfnamefont{H.~R.} \bibnamefont{Krishna-murthy}},
  \bibinfo{author}{\bibfnamefont{J.~W.} \bibnamefont{Wilkins}},
  \bibnamefont{and} \bibinfo{author}{\bibfnamefont{K.~G.}
  \bibnamefont{Wilson}}, \bibinfo{journal}{Phys. Rev. B}
  \textbf{\bibinfo{volume}{21}}, \bibinfo{pages}{1003} (\bibinfo{year}{1980}),
  \urlprefix\url{http://link.aps.org/doi/10.1103/PhysRevB.21.1003}.

\bibitem[{\citenamefont{Bulla et~al.}(2008)\citenamefont{Bulla, Costi, and
  Pruschke}}]{RevModPhys.80.395}
\bibinfo{author}{\bibfnamefont{R.}~\bibnamefont{Bulla}},
  \bibinfo{author}{\bibfnamefont{T.~A.} \bibnamefont{Costi}}, \bibnamefont{and}
  \bibinfo{author}{\bibfnamefont{T.}~\bibnamefont{Pruschke}},
  \bibinfo{journal}{Rev. Mod. Phys.} \textbf{\bibinfo{volume}{80}},
  \bibinfo{pages}{395} (\bibinfo{year}{2008}),
  \urlprefix\url{http://link.aps.org/doi/10.1103/RevModPhys.80.395}.

\bibitem[{\citenamefont{Peters et~al.}(2006)\citenamefont{Peters, Pruschke, and
  Anders}}]{PhysRevB.74.245114}
\bibinfo{author}{\bibfnamefont{R.}~\bibnamefont{Peters}},
  \bibinfo{author}{\bibfnamefont{T.}~\bibnamefont{Pruschke}}, \bibnamefont{and}
  \bibinfo{author}{\bibfnamefont{F.~B.} \bibnamefont{Anders}},
  \bibinfo{journal}{Phys. Rev. B} \textbf{\bibinfo{volume}{74}},
  \bibinfo{pages}{245114} (\bibinfo{year}{2006}),
  \urlprefix\url{http://link.aps.org/doi/10.1103/PhysRevB.74.245114}.

\bibitem[{\citenamefont{Weichselbaum and von
  Delft}(2007{\natexlab{a}})}]{PhysRevLett.99.076402}
\bibinfo{author}{\bibfnamefont{A.}~\bibnamefont{Weichselbaum}}
  \bibnamefont{and} \bibinfo{author}{\bibfnamefont{J.}~\bibnamefont{von
  Delft}}, \bibinfo{journal}{Phys. Rev. Lett.} \textbf{\bibinfo{volume}{99}},
  \bibinfo{pages}{076402} (\bibinfo{year}{2007}{\natexlab{a}}),
  \urlprefix\url{http://link.aps.org/doi/10.1103/PhysRevLett.99.076402}.

\bibitem[{\citenamefont{\v{Z}itko and Pruschke}(2009)}]{PhysRevB.79.085106}
\bibinfo{author}{\bibfnamefont{R.}~\bibnamefont{\v{Z}itko}} \bibnamefont{and}
  \bibinfo{author}{\bibfnamefont{T.}~\bibnamefont{Pruschke}},
  \bibinfo{journal}{Phys. Rev. B} \textbf{\bibinfo{volume}{79}},
  \bibinfo{pages}{085106} (\bibinfo{year}{2009}),
  \urlprefix\url{http://link.aps.org/doi/10.1103/PhysRevB.79.085106}.

\bibitem[{\citenamefont{Bodensiek
  et~al.}(2011{\natexlab{a}})\citenamefont{Bodensiek, {\v{Z}}itko, Peters, and
  Pruschke}}]{bodensiek2011low}
\bibinfo{author}{\bibfnamefont{O.}~\bibnamefont{Bodensiek}},
  \bibinfo{author}{\bibfnamefont{R.}~\bibnamefont{{\v{Z}}itko}},
  \bibinfo{author}{\bibfnamefont{R.}~\bibnamefont{Peters}}, \bibnamefont{and}
  \bibinfo{author}{\bibfnamefont{T.}~\bibnamefont{Pruschke}},
  \bibinfo{journal}{Journal of Physics: Condensed Matter}
  \textbf{\bibinfo{volume}{23}}, \bibinfo{pages}{094212}
  (\bibinfo{year}{2011}{\natexlab{a}}).

\bibitem[{\citenamefont{Peters and
  Pruschke}(2007{\natexlab{a}})}]{PhysRevB.76.245101}
\bibinfo{author}{\bibfnamefont{R.}~\bibnamefont{Peters}} \bibnamefont{and}
  \bibinfo{author}{\bibfnamefont{T.}~\bibnamefont{Pruschke}},
  \bibinfo{journal}{Phys. Rev. B} \textbf{\bibinfo{volume}{76}},
  \bibinfo{pages}{245101} (\bibinfo{year}{2007}{\natexlab{a}}),
  \urlprefix\url{http://link.aps.org/doi/10.1103/PhysRevB.76.245101}.

\bibitem[{\citenamefont{Yoshida et~al.}(1990)\citenamefont{Yoshida, Whitaker,
  and Oliveira}}]{PhysRevB.41.9403}
\bibinfo{author}{\bibfnamefont{M.}~\bibnamefont{Yoshida}},
  \bibinfo{author}{\bibfnamefont{M.~A.} \bibnamefont{Whitaker}},
  \bibnamefont{and} \bibinfo{author}{\bibfnamefont{L.~N.}
  \bibnamefont{Oliveira}}, \bibinfo{journal}{Phys. Rev. B}
  \textbf{\bibinfo{volume}{41}}, \bibinfo{pages}{9403} (\bibinfo{year}{1990}),
  \urlprefix\url{http://link.aps.org/doi/10.1103/PhysRevB.41.9403}.

\bibitem[{\citenamefont{Bulla et~al.}(1998)\citenamefont{Bulla, Hewson, and
  Pruschke}}]{bulla1998numerical}
\bibinfo{author}{\bibfnamefont{R.}~\bibnamefont{Bulla}},
  \bibinfo{author}{\bibfnamefont{A.}~\bibnamefont{Hewson}}, \bibnamefont{and}
  \bibinfo{author}{\bibfnamefont{T.}~\bibnamefont{Pruschke}},
  \bibinfo{journal}{Journal of Physics: Condensed Matter}
  \textbf{\bibinfo{volume}{10}}, \bibinfo{pages}{8365} (\bibinfo{year}{1998}).

\bibitem[{\citenamefont{Weichselbaum and von
  Delft}(2007{\natexlab{b}})}]{weichselbaum2007}
\bibinfo{author}{\bibfnamefont{A.}~\bibnamefont{Weichselbaum}}
  \bibnamefont{and} \bibinfo{author}{\bibfnamefont{J.}~\bibnamefont{von
  Delft}}, \bibinfo{journal}{Phys. Rev. Lett.} \textbf{\bibinfo{volume}{99}},
  \bibinfo{pages}{076402} (\bibinfo{year}{2007}{\natexlab{b}}).

\bibitem[{\citenamefont{\v{Z}itko}(2009)}]{broyden}
\bibinfo{author}{\bibfnamefont{R.}~\bibnamefont{\v{Z}itko}},
  \bibinfo{journal}{Phys. Rev. B} \textbf{\bibinfo{volume}{80}},
  \bibinfo{pages}{125125} (\bibinfo{year}{2009}).

\bibitem[{\citenamefont{Taranto et~al.}(2012)\citenamefont{Taranto,
  Sangiovanni, Held, Capone, Georges, and Toschi}}]{PhysRevB.85.085124}
\bibinfo{author}{\bibfnamefont{C.}~\bibnamefont{Taranto}},
  \bibinfo{author}{\bibfnamefont{G.}~\bibnamefont{Sangiovanni}},
  \bibinfo{author}{\bibfnamefont{K.}~\bibnamefont{Held}},
  \bibinfo{author}{\bibfnamefont{M.}~\bibnamefont{Capone}},
  \bibinfo{author}{\bibfnamefont{A.}~\bibnamefont{Georges}}, \bibnamefont{and}
  \bibinfo{author}{\bibfnamefont{A.}~\bibnamefont{Toschi}},
  \bibinfo{journal}{Phys. Rev. B} \textbf{\bibinfo{volume}{85}},
  \bibinfo{pages}{085124} (\bibinfo{year}{2012}),
  \urlprefix\url{http://link.aps.org/doi/10.1103/PhysRevB.85.085124}.

\bibitem[{\citenamefont{Hewson}(1993)}]{hewson}
\bibinfo{author}{\bibfnamefont{A.~C.} \bibnamefont{Hewson}},
  \emph{\bibinfo{title}{The Kondo Problem to Heavy-Fermions}}
  (\bibinfo{publisher}{Cambridge University Press, Cambridge},
  \bibinfo{year}{1993}).

\bibitem[{\citenamefont{Rozenberg et~al.}(1996)\citenamefont{Rozenberg,
  Kotliar, and Kajueter}}]{rozenberg1996}
\bibinfo{author}{\bibfnamefont{M.~J.} \bibnamefont{Rozenberg}},
  \bibinfo{author}{\bibfnamefont{G.}~\bibnamefont{Kotliar}}, \bibnamefont{and}
  \bibinfo{author}{\bibfnamefont{H.}~\bibnamefont{Kajueter}},
  \bibinfo{journal}{Phys. Rev. B} \textbf{\bibinfo{volume}{54}},
  \bibinfo{pages}{8452} (\bibinfo{year}{1996}).

\bibitem[{\citenamefont{Georges
  et~al.}(1996{\natexlab{b}})\citenamefont{Georges, Kotliar, Krauth, and
  Rozenberg}}]{georges1996}
\bibinfo{author}{\bibfnamefont{A.}~\bibnamefont{Georges}},
  \bibinfo{author}{\bibfnamefont{G.}~\bibnamefont{Kotliar}},
  \bibinfo{author}{\bibfnamefont{W.}~\bibnamefont{Krauth}}, \bibnamefont{and}
  \bibinfo{author}{\bibfnamefont{M.~J.} \bibnamefont{Rozenberg}},
  \bibinfo{journal}{Rev. Mod. Phys.} \textbf{\bibinfo{volume}{68}},
  \bibinfo{pages}{13} (\bibinfo{year}{1996}{\natexlab{b}}).

\bibitem[{\citenamefont{Rozenberg}(1995)}]{rozenberg1995}
\bibinfo{author}{\bibfnamefont{M.~J.} \bibnamefont{Rozenberg}},
  \bibinfo{journal}{Phys. Rev. B} \textbf{\bibinfo{volume}{52}},
  \bibinfo{pages}{7369} (\bibinfo{year}{1995}).

\bibitem[{\citenamefont{Capponi and Assaad}(2001)}]{capponi2001}
\bibinfo{author}{\bibfnamefont{S.}~\bibnamefont{Capponi}} \bibnamefont{and}
  \bibinfo{author}{\bibfnamefont{F.~F.} \bibnamefont{Assaad}},
  \bibinfo{journal}{Phys. Rev. B} \textbf{\bibinfo{volume}{63}},
  \bibinfo{pages}{155114} (\bibinfo{year}{2001}).

\bibitem[{\citenamefont{Zitzler et~al.}(2002)\citenamefont{Zitzler, Pruschke,
  and Bulla}}]{zitzler2002}
\bibinfo{author}{\bibfnamefont{R.}~\bibnamefont{Zitzler}},
  \bibinfo{author}{\bibfnamefont{T.}~\bibnamefont{Pruschke}}, \bibnamefont{and}
  \bibinfo{author}{\bibfnamefont{R.}~\bibnamefont{Bulla}},
  \bibinfo{journal}{Eur. Phys. J. B} \textbf{\bibinfo{volume}{27}},
  \bibinfo{pages}{473} (\bibinfo{year}{2002}).

\bibitem[{\citenamefont{Bodensiek
  et~al.}(2011{\natexlab{b}})\citenamefont{Bodensiek, \v{Z}itko, Peters, and
  Pruschke}}]{oliver}
\bibinfo{author}{\bibfnamefont{O.}~\bibnamefont{Bodensiek}},
  \bibinfo{author}{\bibfnamefont{R.}~\bibnamefont{\v{Z}itko}},
  \bibinfo{author}{\bibfnamefont{R.}~\bibnamefont{Peters}}, \bibnamefont{and}
  \bibinfo{author}{\bibfnamefont{T.}~\bibnamefont{Pruschke}},
  \bibinfo{journal}{J. Phys.: Condens. Matter} p. \bibinfo{pages}{094212}
  (\bibinfo{year}{2011}{\natexlab{b}}).

\bibitem[{\citenamefont{Peters and
  Pruschke}(2007{\natexlab{b}})}]{peters2007magnetic}
\bibinfo{author}{\bibfnamefont{R.}~\bibnamefont{Peters}} \bibnamefont{and}
  \bibinfo{author}{\bibfnamefont{T.}~\bibnamefont{Pruschke}},
  \bibinfo{journal}{Phys. Rev. B} \textbf{\bibinfo{volume}{76}},
  \bibinfo{pages}{245101} (\bibinfo{year}{2007}{\natexlab{b}}).

\bibitem[{\citenamefont{Peters et~al.}(2011)\citenamefont{Peters, Kawakami, and
  Pruschke}}]{Peters:2011iq}
\bibinfo{author}{\bibfnamefont{R.}~\bibnamefont{Peters}},
  \bibinfo{author}{\bibfnamefont{N.}~\bibnamefont{Kawakami}}, \bibnamefont{and}
  \bibinfo{author}{\bibfnamefont{T.}~\bibnamefont{Pruschke}},
  \bibinfo{journal}{Journal of Physics: Conference Series}
  \textbf{\bibinfo{volume}{320}}, \bibinfo{pages}{012057}
  (\bibinfo{year}{2011}).

\bibitem[{\citenamefont{Beach et~al.}(2004)\citenamefont{Beach, Lee, and
  Monthoux}}]{PhysRevLett.92.026401}
\bibinfo{author}{\bibfnamefont{K.~S.~D.} \bibnamefont{Beach}},
  \bibinfo{author}{\bibfnamefont{P.~A.} \bibnamefont{Lee}}, \bibnamefont{and}
  \bibinfo{author}{\bibfnamefont{P.}~\bibnamefont{Monthoux}},
  \bibinfo{journal}{Phys. Rev. Lett.} \textbf{\bibinfo{volume}{92}},
  \bibinfo{pages}{026401} (\bibinfo{year}{2004}),
  \urlprefix\url{http://link.aps.org/doi/10.1103/PhysRevLett.92.026401}.

\bibitem[{\citenamefont{Ohashi et~al.}(2004)\citenamefont{Ohashi, Koga, Suga,
  and Kawakami}}]{PhysRevB.70.245104}
\bibinfo{author}{\bibfnamefont{T.}~\bibnamefont{Ohashi}},
  \bibinfo{author}{\bibfnamefont{A.}~\bibnamefont{Koga}},
  \bibinfo{author}{\bibfnamefont{S.-i.} \bibnamefont{Suga}}, \bibnamefont{and}
  \bibinfo{author}{\bibfnamefont{N.}~\bibnamefont{Kawakami}},
  \bibinfo{journal}{Phys. Rev. B} \textbf{\bibinfo{volume}{70}},
  \bibinfo{pages}{245104} (\bibinfo{year}{2004}),
  \urlprefix\url{http://link.aps.org/doi/10.1103/PhysRevB.70.245104}.

\bibitem[{\citenamefont{Bodensiek et~al.}(2013)\citenamefont{Bodensiek,
  \v{Z}itko, Vojta, Jarrell, and Pruschke}}]{PhysRevLett.110.146406}
\bibinfo{author}{\bibfnamefont{O.}~\bibnamefont{Bodensiek}},
  \bibinfo{author}{\bibfnamefont{R.}~\bibnamefont{\v{Z}itko}},
  \bibinfo{author}{\bibfnamefont{M.}~\bibnamefont{Vojta}},
  \bibinfo{author}{\bibfnamefont{M.}~\bibnamefont{Jarrell}}, \bibnamefont{and}
  \bibinfo{author}{\bibfnamefont{T.}~\bibnamefont{Pruschke}},
  \bibinfo{journal}{Phys. Rev. Lett.} \textbf{\bibinfo{volume}{110}},
  \bibinfo{pages}{146406} (\bibinfo{year}{2013}),
  \urlprefix\url{http://link.aps.org/doi/10.1103/PhysRevLett.110.146406}.

\bibitem[{\citenamefont{Tanaskovi{\'c}
  et~al.}(2011)\citenamefont{Tanaskovi{\'c}, Haule, Kotliar, and
  Dobrosavljevi{\'c}}}]{Tanaskovic:2011jk}
\bibinfo{author}{\bibfnamefont{D.}~\bibnamefont{Tanaskovi{\'c}}},
  \bibinfo{author}{\bibfnamefont{K.}~\bibnamefont{Haule}},
  \bibinfo{author}{\bibfnamefont{G.}~\bibnamefont{Kotliar}}, \bibnamefont{and}
  \bibinfo{author}{\bibfnamefont{V.}~\bibnamefont{Dobrosavljevi{\'c}}},
  \bibinfo{journal}{Physical Review B} \textbf{\bibinfo{volume}{84}},
  \bibinfo{pages}{115105} (\bibinfo{year}{2011}).

\end{thebibliography}

\end{document}